\documentclass{aa}

\usepackage{txfonts}
\usepackage{graphicx}
\usepackage{natbib}
\bibpunct{(}{)}{;}{a}{}{,}

\begin{document}

\title{J-PLUS: Morphological star/galaxy classification by PDF analysis}

\author{C.~L\'opez-Sanjuan\inst{1}
\and H.~V\'azquez Rami\'o\inst{1}
\and J.~Varela\inst{1}
\and D.~Spinoso\inst{1}
\and R.~E.~Angulo\inst{1}
\and D.~Muniesa\inst{1}
\and K.~Viironen\inst{1}
\and D.~Crist\'obal-Hornillos\inst{1}
\and A.~J.~Cenarro\inst{1}
\and A.~Ederoclite\inst{1}
\and A.~Mar\'{\i}n-Franch\inst{1}
\and M.~Moles\inst{1}
\and B.~Ascaso\inst{2}  
\and S.~Bonoli\inst{1}
\and A.~L.~Chies-Santos\inst{3}
\and P.~R.~T.~Coelho\inst{4}
\and M.~V.~Costa-Duarte\inst{4}
\and A.~Cortesi\inst{4}
\and L.~A.~Díaz-García\inst{1}
\and R.~A.~Dupke\inst{5,6}
\and L.~Galbany\inst{7}
\and C.~Hern\'andez-Monteagudo\inst{1}
\and R.~Logro\~no-Garc\'{\i}a\inst{1}
\and A.~Molino\inst{4}
\and A.~Orsi\inst{1}
\and V.~M.~Placco\inst{8}
\and L.~Sampedro\inst{4}
\and I.~San Roman\inst{1}
\and G.~Vilella-Rojo\inst{1}
\and D.~D.~Whitten\inst{8}
\and C.~L.~Mendes de Oliveira\inst{4}
\and L.~Sodr\'e Jr.\inst{4}
}

\institute{Centro de Estudios de F\'{\i}sica del Cosmos de Arag\'on, Plaza San Juan 1, 44001 Teruel, Spain\\\email{clsj@cefca.es}   
	\and
APC, AstroParticule et Cosmologie, Universit\`e Paris Diderot, CNRS/IN2P3, CEA/lrfu, Observatoire de Paris, Sorbonne Paris Cit\'e, 10, rue Alice Domon et L\'eonie Duquet, 75205 Paris Cedex 13, France
	\and
Departamento de Astronomia, Instituto de F\'{\i}sica, Universidade Federal do Rio Grande do Sul, Porto Alegre, R.S, Brazil
	\and
Instituto de Astronomia, Geof\'{\i}sica e Ci\^encias Atmosf\'ericas, Universidade de S\~ao Paulo, 05508-090 S\~ao Paulo, Brazil
	\and
Observat\'orio Nacional/MCTIC, Rua Gen. Jos\'e Cristino 77, 20921-400, Rio de Janeiro, Brazil
	\and
Dept. of Astronomy, University of Michigan, Ann Arbor, MI 48109-1107, USA
	\and
PITT PACC, Department of Physics and Astronomy, University of Pittsburgh, Pittsburgh, PA 15260, USA
	\and
Department of Physics and JINA Center for the Evolution of the Elements, University of Notre Dame, Notre Dame, IN 46556, USA
}

\date{Submitted December 2017}

\abstract
{}
{Our goal is to morphologically classify the sources identified in the images of the J-PLUS early data release (EDR) into compact (stars) or extended (galaxies) using a suited Bayesian classifier.}
{J-PLUS sources exhibit two distinct populations in the $r$-band magnitude vs. concentration plane, corresponding to compact and extended sources. We modelled the two-population distribution with a skewed Gaussian for compact objects and a log-normal function for the extended ones. The derived model and the number density prior based on J-PLUS EDR data were used to estimate the Bayesian probability of a source to be star or galaxy. This procedure was applied pointing-by-pointing to account for varying observing conditions and sky position. Finally, we combined the morphological information from $g$, $r$, and $i$ broad bands in order to improve the classification of low signal-to-noise sources.}
{The derived probabilities are used to compute the pointing-by-pointing number counts of stars and galaxies. The former increases as we approach to the Milky Way disk, and the latter are similar across the probed area. The comparison with SDSS in the common regions is satisfactory up to $r \sim 21$, with consistent numbers of stars and galaxies, and consistent distributions in concentration and $(g-i)$ colour spaces.}
{We implement a morphological star/galaxy classifier based on PDF analysis, providing meaningful probabilities for J-PLUS sources to one magnitude deeper ($r \sim 21$) than a classical boolean classification. These probabilities are suited for the statistical study of 150k stars and 101k galaxies with $15 < r \leq 21$ present in the $31.7$ deg$^2$ of the J-PLUS EDR. In a future version of the classifier, we will include J-PLUS colour information from twelve photometric bands.}


\titlerunning{J-PLUS. Morphological star/galaxy classification by PDF analysis}

\authorrunning{L\'opez-Sanjuan et al.}

\maketitle

\section{Introduction}\label{intro}
The study of Milky Way stars and the understanding of extragalactic sources have greatly benefited from large ($\gtrsim 5000$~deg$^2$) and systematic photometric surveys, such as the POSS-II (second Palomar Observatory Sky Survey, three optical $gri$ broad bands; \citealt{poss2}), SDSS (Sloan Digital Sky Survey, five optical $ugriz$ broad bands; \citealt{sdssdr7}), or VHS (VISTA Hemisphere Survey, three near-infrared $HJK_{\rm s}$ bands; \citealt{vhs}); and will move forward in the following years with next generation surveys such as DES (Dark Energy Survey, five optical $ugriz$ broad bands; \citealt{des}), UHS (UKIRT Hemisphere Survey, two near-infrared $JK_{\rm s}$ bands; \citealt{uhs}), {\it Euclid} (three near-infrared $YHJ$ broad bands; \citealt{euclid}), LSST (Large Synoptic Survey Telescope, six optical $ugrizY$ broad bands; \citealt{lsst}), J-PLUS\footnote{\url{j-plus.es}} (Javalambre Local Universe Survey, five $ugriz$ broad + seven narrow optical bands; \citealt{cenarro18}), S-PLUS (Southern Local Universe Survey, the southern counterpart of J-PLUS; \citealt{splus}), and J-PAS\footnote{\url{j-pas.org}} (Javalambre Physics of the accelerating universe Astrophysical Survey, 56 narrow optical bands; \citealt{jpas}). 

\begin{table*} 
\caption{J-PLUS filter system and limiting magnitudes (3$\sigma$, 3$^{\prime\prime}$ aperture) of J-PLUS and the EDR.} 
\label{tab:JPLUS_filters}
\centering 
	\begin{tabular}{l c c c c l }
	\hline\hline\rule{0pt}{3ex} 
	Name 	& Central Wavelength 	& FWHM 	& $m_{\rm lim}^{\rm J-PLUS}$    & $m_{\rm lim}^{\rm EDR}$   & Comments\\\rule{0pt}{2ex} 
	       	&   [nm]             	& [nm] 	&  [AB]		&  [AB]  &         \\
	\hline\rule{0pt}{2ex}
	$u$	 	&348.5 	&50.8		& 21.00	&	21.6	& In common with J-PAS\\ 
	$J0378$ 	&378.5 	&16.8 		& 21.00 & 	21.5	& [OII]; in common with J-PAS\\ 
	$J0395$ 	&395.0	&10.0		& 21.00 & 	21.4	& Ca H$+$K\\ 
	$J0410$ 	&410.0	&20.0		& 21.25 & 	21.5	& H$_\delta$\\ 
	$J0430$		&430.0 	&20.0		& 21.25 & 	21.5	& G-band\\ 
	$g$		&480.3 	&140.9		& 22.00 & 	22.2	& SDSS\\ 
	$J0515$ 	&515.0	&20.0		& 21.25 & 	21.4	& Mg$b$ Triplet\\ 
	$r$ 		&625.4	&138.8		& 22.00 & 	21.9	& SDSS\\ 
	$J0660$ 	&660.0 	&13.8		& 21.25 & 	21.3	& H$\alpha$; in common with J-PAS\\ 
	$i$		&766.8 	&153.5		& 21.75 & 	20.8	& SDSS\\ 
	$J0861$		&861.0 	&40.0		& 20.50 & 	20.8	& Ca Triplet\\ 
	$z$ 		&911.4	&140.9		& 20.75 & 	20.5	& SDSS\\ 
	\hline 
\end{tabular}
\end{table*}

Because of their photometric nature, the above surveys image all the astronomical sources down to a limiting magnitude, without any pre-selection. Thus, one of the main steps in the analysis of these rich photometric datasets is the classification of the observed sources into stars, galaxies, or other objects of interest (e.g. QSOs, planetary nebula, etc.) 

The star/galaxy classification of astronomical sources is usually performed following two complementary approaches, based on morphology and colours, respectively. Morphological classifiers use the different concentration properties of stars (point-like sources) and galaxies (extended sources) to separate them \citep[e.g.][]{kron80,reid96,sextractor,odewahn04,vasconcellos11}, and colour classifiers take advantage of the different location of stars, galaxies, and QSOs in colour-colour diagrams, using one or more of them to perform the classification \citep[e.g.][]{huang97,elston06,baldry10,saglia12,malek13}. The combination of both approaches is also possible, maximising the use of the available information \citep[e.g.][]{molino13, kim15, soumagnac15, kim17}. 

From the technical point of view, there are several ways to address the source classification problem. We highlight those based on the modelling of the observed spectral energy distribution \citep[SED; e.g.][]{combo17,robin07,saglia12,preethi14}, the application of machine learning codes and neural networks \citep[e.g.][]{cortiglioni01,ball06sg,malek13,miller17}, or the inclusion of prior information from a Bayesian framework \citep[e.g.][]{sebok79,scranton02,henrion11,molino13}.

In the present paper, we implement a morphological Bayesian classifier that accounts for the observational errors in the concentration measurements and applies a magnitude prior to deal with the larger number of galaxies expected at fainter magnitudes. The classifier was developed in the context to the PROFUSE\footnote{\url{profuse.cefca.es}} project, that uses PRObabiliy Functions for Unbiased Statistical Estimations in multi-filter surveys, and was applied to classify the sources of the J-PLUS early data release (EDR) dataset. J-PLUS is a multi-filter survey of thousands square degrees in the northern hemisphere with a set of twelve (five broad + seven narrow) optical filters particularly defined to provide reliable SEDs of Milky Way stars and nearby galaxies, and to photometrically calibrate J-PAS. The J-PLUS survey and its EDR are fully described in \citet{cenarro18}.

The paper is organized as follows. In Sec.~\ref{data}, we present the J-PLUS EDR dataset. The Bayesian classifier and its application to J-PLUS EDR data are described in Sec.~\ref{psjplus}. We test the performance of our classification in Sect.~\ref{test}, and present the star and galaxy number counts of the J-PLUS EDR in Sec.~\ref{counts}. We discuss our results in Sec.~\ref{discussion}, presenting our conclusions in Sec.~\ref{conclusions}. Magnitudes are given in the AB system \citep{oke83}.

\section{J-PLUS early data release}\label{data}
As previously mentioned, J-PLUS is a photometric survey of several thousand square degrees that is being conducted from the Observatorio Astrof\'{\i}sico de Javalambre (OAJ, Teruel, Spain; \citealt{oaj}), using the 83\,cm Javalambre Auxiliary Survey Telescope (JAST/T80) and T80Cam, a panoramic camera that provides a $2\deg^2$ field of view (FoV) with a pixel scale of 0.55$^{\prime\prime}$. The J-PLUS filter system, composed by twelve bands, is summarised in Table~\ref{tab:JPLUS_filters}. The J-PLUS observational strategy, image reduction, photometric calibration, and main scientific goals are presented in \citet{cenarro18}.

The 18 pointings that compose the J-PLUS EDR are representative of the 205 J-PLUS pointings gathered in similar production stage (i.e. reduction and calibration) by September 2017. The limiting magnitudes targeted by J-PLUS and those of the EDR are presented in Table~\ref{tab:JPLUS_filters} for reference. The median point spread function (PSF) full-with half maximum (FWHM) in the EDR $r-$band images is 1.1$^{\prime\prime}$. Source detection was done in the $r-$band using \texttt{SExtractor} \citep{sextractor}, and the flux measured in the twelve J-PLUS bands at the position of the detected sources using its dual mode capability. The EDR is publicly available at the J-PLUS web site\footnote{\url{j-plus.es/datareleases/early_data_release}}.

In addition to the present paper, the J-PLUS EDR and science verification data have been used to refine the membership in the nearby galaxy clusters A2589 \& A2593 \citep{molino18}, analyse the globular cluster M15 \citep{bonatto18}, and study the H$\alpha$ emission \citep{logronho18} and the stellar populations \citep{sanroman18} of several local galaxies.


\subsection{Masking and final J-PLUS EDR area}\label{mask}
To perform our study, we were restricted to the high signal-to-noise areas covered by the J-PLUS EDR. We defined the best area for each pointing as those pixels with relative exposure higher than 0.7 with respect to the maximum in the $r-$band detection image\footnote{J-PLUS pointings comprise three exposures. Hence, the 0.7 exposure condition selects areas covered by the three acquired images at each position}. This avoids the borders of the images and provides a well defined area of $\sim 2$ deg$^2$ per pointing, with 36 deg$^2$ covered by the J-PLUS EDR without accounting for the overlaps between adjacent pointings.

We masked the areas surrounding bright stars that are affected by spikes and luminous haloes, compromising the photometry of the sources. This was done by searching for the Tycho2 stars \citep{tycho2} and masking a circular area centred in the position of the star with radius
\begin{equation}
0.075\,(12 - V_{\rm T})^{2.4} + 1.5\ \ {\rm [arcmin]},
\end{equation}
where $V_{\rm T}$ is the $V-$band magnitude of the star in the Tycho photometric system. The masking was only applied to stars brighter than $V_{\rm T} = 12$. The circular area was estimated empirically to minimise the impact of bright stars, improving the reliability of the photometry. The area masked because of bright stars is 4 deg$^2$, 11\% of the initial J-PLUS EDR area.

Finally, we masked the areas affected by reflections and artefacts. The artefacts mask was defined by visual inspection of the J-PLUS EDR images. The area masked because of artefacts is only 0.02 deg$^2$, less than 0.1\% of the initial area.

The final high-quality area of the J-PLUS EDR after masking is 31.98 deg$^2$. Accounting for the overlapping areas between adjacent pointings, the effective area is reduced to 31.70 deg$^2$. The former area refers to pointing-by-pointing studies that include duplicated sources observed in more than one pointing, and the latter to the J-PLUS EDR studies that include only unique sources. The present paper performs a pointing-by-pointing morphological classification to cope with the varying conditions of the observations (Sect.~\ref{slocus}). The masks are available at the public J-PLUS data base.

The J-PLUS EDR comprises a total of 292k unique detections in the $r$-band images with $15 < r \leq 21$, which is reduced to 251k (84\%) after masking.

\begin{figure}[t]
\centering
\resizebox{\hsize}{!}{\includegraphics{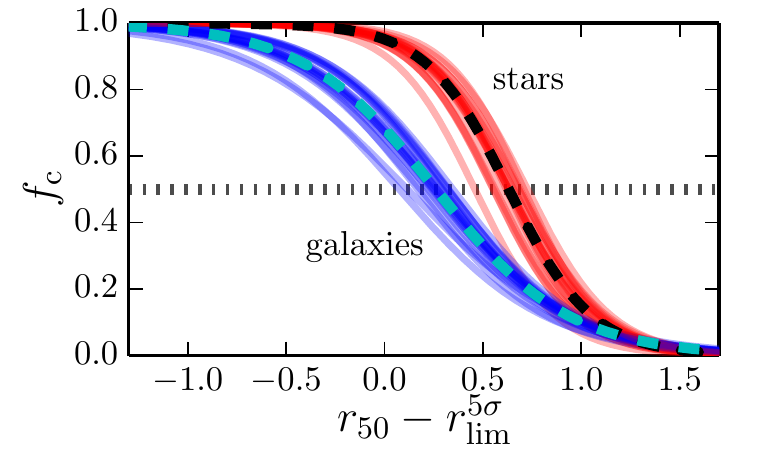}}
\caption{Completeness curves referenced to the 5$\sigma$ detection in the J-PLUS $r-$band magnitude ($r_{\rm lim}^{5\sigma}$) for stars (red lines) and galaxies (blue lines) in each of the J-PLUS EDR pointings. The dashed lines show the median relations, with $r_{50,{\rm s}} = 0.63 + r_{\rm lim}^{5\sigma}$ and $\kappa_{\rm s} = 4.7$ for stars (black), and $r_{50,{\rm g}} = 0.26 + r_{\rm lim}^{5\sigma}$ and $\kappa_{\rm g} = 3.0$ for galaxies (cyan).}
\label{fc_edr}
\end{figure}

\subsection{Detection completeness of stars and galaxies}\label{completeness}
We used SDSS photometry as a reference to estimate the star and galaxy completeness of the J-PLUS EDR pointings in the $r$ band. We compared the number of stars and galaxies in the general SDSS catalogue with those from the sources detected in the J-PLUS $r-$band images. To avoid biases related with the lower signal-to-noise (${\rm S/N}$) of J-PLUS images close to the limiting magnitude, we used the star/galaxy classification and the model magnitudes from SDSS, that is typically one magnitude deeper than J-PLUS images in the $r$ band.

The measured $r-$band completeness is then parametrised with the function
\begin{equation}
f_{\rm c}\,(r) = 1 - \frac{1}{1 + {\rm exp}[ -\kappa\,(r - r_{50}) ]} ,
\end{equation}
where $r_{50}$ is the magnitude at 50\% completeness and $\kappa$ the decay rate of the completeness. The obtained curves in each pointing for both stars and galaxies with respect to the 5$\sigma$ detection in the J-PLUS $r-$band \texttt{AUTO} magnitude ($r_{\rm lim}^{5\sigma}$) are presented in Fig.~\ref{fc_edr}. We find evident similarities among the completeness curves of different pointings after normalizing them to the depth of the corresponding image, with stars being detected to fainter magnitudes and presenting a steeper decay rate. Hence, for simplicity, in the following we assume a common decay rate $\kappa_{\rm s} = 4.7$ for stars and $\kappa_{\rm g} = 3.0$ for galaxies, and define the 50\% completeness magnitude of each EDR pointing as $r_{50,{\rm s}} = 0.63 + r_{\rm lim}^{5\sigma}$ for stars and $r_{50,{\rm g}} = 0.26 + r_{\rm lim}^{5\sigma}$ for galaxies. The median 50\% completeness of the J-PLUS EDR is reached at $r = 21.51$ for stars and $r = 21.10$ for galaxies, magnitudes that change to $r = 20.88$ and $r = 20.12$, respectively, for 95\% completeness. 

The above procedure is valid for J-PLUS pointings in shared areas with SDSS and only for the broad bands in common. We plan to estimate the completeness for stars and galaxies in all the J-PLUS filters and pointings by injecting realistic fake sources and studying their recovery rate (see \citealt{molino18}, for details about the procedure).

\section{Morphological classification by PDF analysis}\label{psjplus}
Given an image with a set of properties (i.e. pixel size, PSF FWHM, photometric depth, position in the sky, etc.), we can classify the detected sources in different astronomical categories using their morphological properties. The most basic distinction is between compact (e.g. stars and QSOs) and extended (e.g. galaxies and nebulae) sources.

To perform a morphological separation, we need to define a compactness indicator for each detection in our images. Such indicator can be based on the comparison between properly selected magnitudes \citep[e.g.][]{yasuda01}. In our case, we defined the concentration $c_r$ by the difference between the $r$-band magnitudes estimated at $1.5^{\prime\prime}$ ($r_{1.5^{\prime\prime}}$) and $3.0^{\prime\prime}$ ($r_{3.0^{\prime\prime}}$) diameter apertures,
\begin{equation}
c_{r} = r_{1.5^{\prime\prime}} - r_{3.0^{\prime\prime}}.
\end{equation}
The concentration $c_r$ is smaller for more compact, point-like sources, and larger for extended, less concentrated objects. We note that our definition approximates the light growth curve of the sources with only two points, that in this case roughly corresponds to one and two PSF FWHM in J-PLUS EDR $r-$band images. We checked that this combination of aperture magnitudes provides the best results for our purpose.

\begin{figure}[t]
\centering
\resizebox{\hsize}{!}{\includegraphics{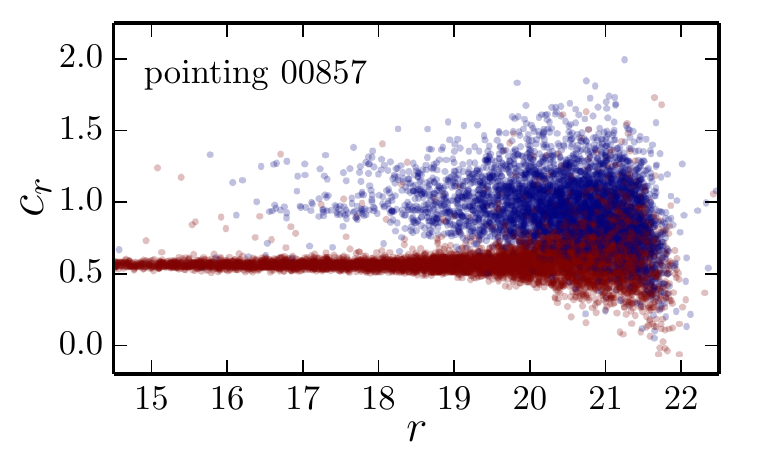}}
\resizebox{\hsize}{!}{\includegraphics{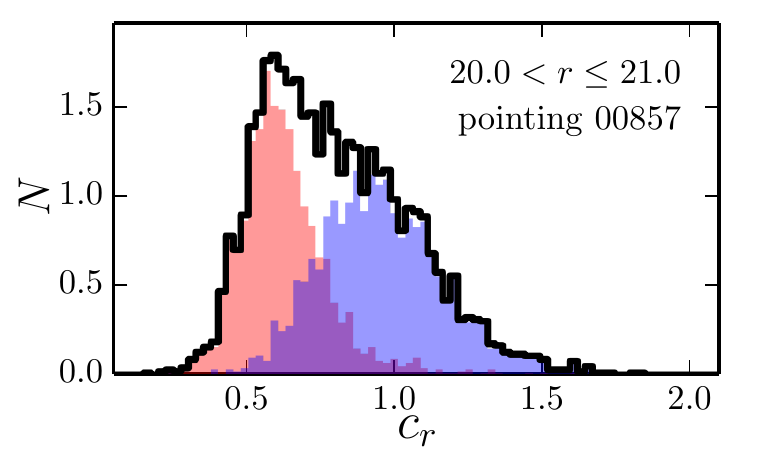}}
\caption{Concentration vs. $r-$band magnitude for the J-PLUS EDR pointing 00857 ({\it top panel}), and distribution of $20 < r \leq 21$ sources in concentration space (black histogram in the {\it bottom panel}). In both panels, sources classified as stars by SDSS are in red, and those classified as galaxies in blue.}
\label{magcon_sdss}
\end{figure}

The distribution of the concentration parameter with respect to the $r-$band \texttt{AUTO} magnitude, a proxy for the total magnitude of the sources, for J-PLUS pointing 00857 is presented in the {\it top panel} of Fig.~\ref{magcon_sdss}. We note that this pointing has a PSF FWHM $=1.1^{\prime\prime}$, similar to the median of the J-PLUS EDR, and we use it as a representative example throughout this paper. Two populations are evident in the concentration vs. magnitude plane: a tight sequence of compact sources at $c_r \sim 0.55$ and a cloud of extended ones at $c_r \sim 1$, with concentration spanning a range between 0.2 and 2.0. Both populations are well separated at $r \lesssim 19.5$, and merge at fainter magnitudes. The vast majority of compact sources in our magnitude range ($r \leq 21$) are Milky Way stars, whereas nearly all extended sources are galaxies. Hence, we denote compact sources as stars and extended sources as galaxies in the following. 

The SDSS provides a morphological classification based on the comparison between a PSF magnitude, computed assuming a compact PSF-like light profile, and a model magnitude, computed assuming a combination of S\'ersic profiles. The model and the PSF magnitudes are similar for compact, star-like sources, and different for extended sources \citep{yasuda01,strauss02}. The SDSS classification is reliable up to $r \sim 21$, and it is used as a reference throughout the present paper. We used SDSS classification to colour code the J-PLUS sources in Fig~\ref{magcon_sdss}. As expected, sources classified as compact in SDSS are located in the sequence of lower concentrations, and the extended ones populate the cloud at larger $c_{r}$ values. Because of the uncertainties in our measurements, both distributions broaden at $r \gtrsim 19$, making it difficult to define a clear separation between stars and galaxies at fainter magnitudes. This is further illustrated in the {\it bottom panel} of Fig.~\ref{magcon_sdss}, where the concentration distribution of stars and galaxies following SDSS classification at $20 < r \leq 21$ is shown. Both distributions overlap, with their tails crossing at $c_{r} \sim 0.75$, and a significant number of compact sources have larger $c_{r}$ values than some extended sources. Hence, it seems unfeasible to define a criteria to classify J-PLUS EDR sources with $r > 19.5$ based on the parameter $c_{r}$ alone. 

We analysed J-PLUS EDR sources with a Bayesian star/galaxy classifier \citep{sebok79,scranton02,henrion11,molino13}, providing the probability of each source to be a star or a galaxy. This method was designed to provide meaningful probabilities for low signal-to-noise sources, pushing the study of stars and galaxies in J-PLUS EDR to magnitudes fainter than $r \sim 19.5$. We used SDSS classification to test the quality of our results is Sect~\ref{test}, but we stress that the derived classification is based solely on J-PLUS information.

\begin{figure*}[t]
\centering
\resizebox{0.49\hsize}{!}{\includegraphics{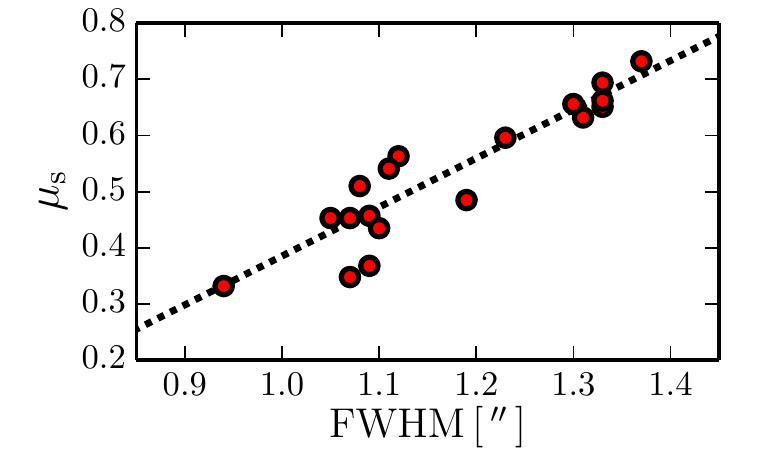}}
\resizebox{0.49\hsize}{!}{\includegraphics{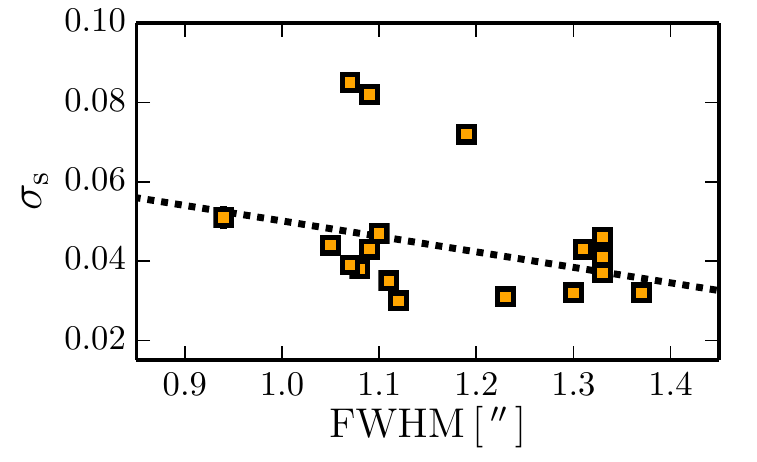}}
\caption{Variation of the parameters defining the stellar locus (position $\mu_{\rm s}$ in the {\it left panel} and dispersion $\sigma_{\rm s}$ in the {\it right panel}) as a function of the PSF FWHM of the J-PLUS EDR $r-$band image. The dotted line in each panel is the error-weighted linear fit to the points.}
\label{slocus_fig}
\end{figure*}

\begin{figure*}[t]
\centering
\resizebox{0.49\hsize}{!}{\includegraphics{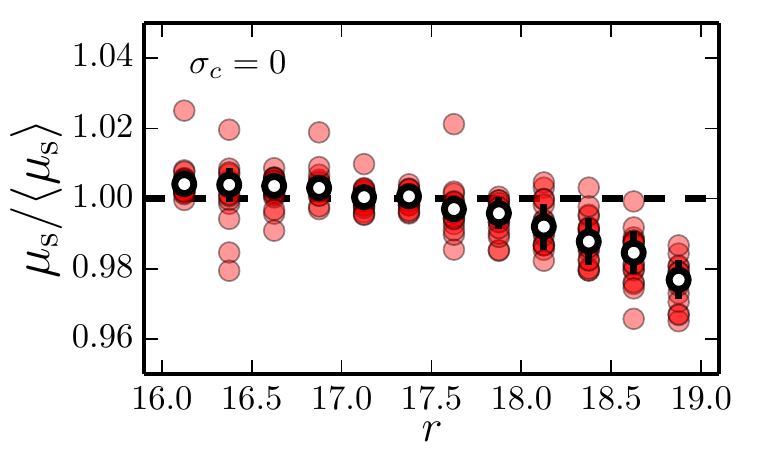}}
\resizebox{0.49\hsize}{!}{\includegraphics{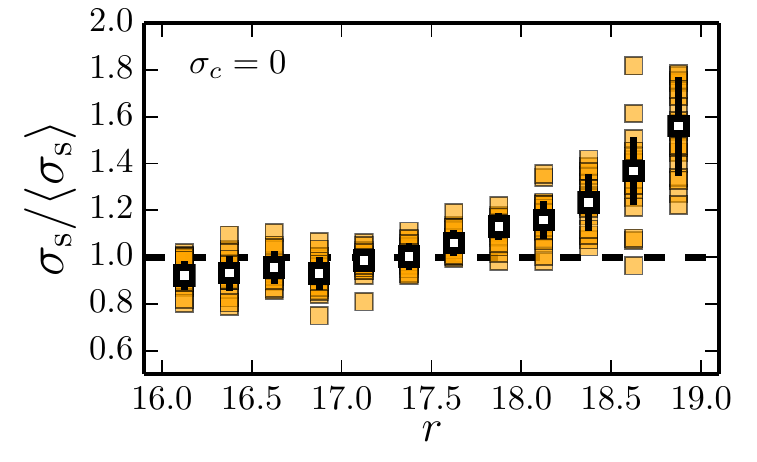}}\\
\resizebox{0.49\hsize}{!}{\includegraphics{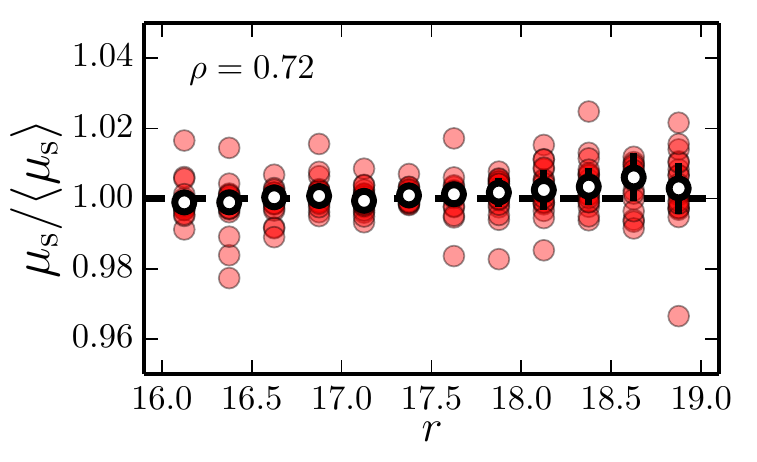}}
\resizebox{0.49\hsize}{!}{\includegraphics{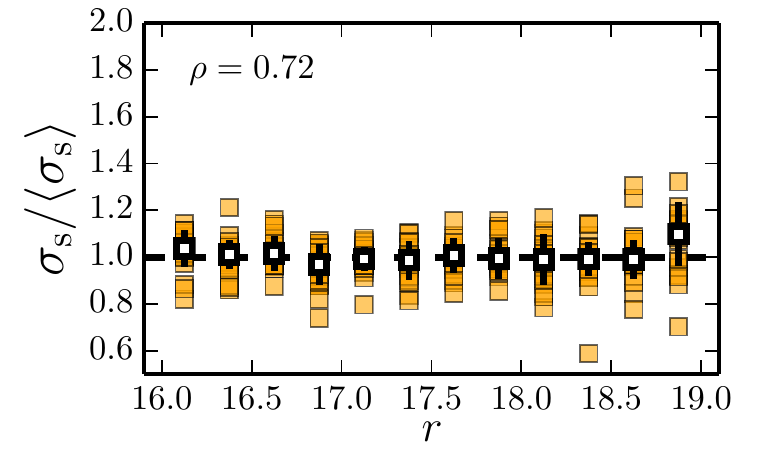}}
\caption{Variation of the normalized stellar locus position (circles in the {\it left panels}) and dispersion (squares in the {\it right panels}) with $r-$band magnitude. The coloured symbols mark the measurements in each J-PLUS EDR pointing. The white symbols show the median and the dispersion of the pointing-based values. The dashed lines mark identity. The {\it top panels} show the results assuming no concentration uncertainties ($\sigma_c = 0$), and the {\it bottom panels} with the uncertainties, computed assuming a covariance $\rho = 0.72$, properly accounted for.}
\label{slocus_mag}
\end{figure*}

\subsection{Probability distribution function of compact and extended sources}
In this section, we present the mathematical formalism and the main ingredients of the Bayesian star/galaxy classifier. Further details can be found in \citet{scranton02} and \citet{henrion11}, so we focus on its application to J-PLUS EDR dataset.
 
We want to estimate the probability distribution function (PDF) in type $t$ space for stars (${\rm s}$) and galaxies (${\rm g}$), $\vec{t} = ({\rm s}, {\rm g})$, given a measured concentration $c$ and its error $\sigma_c$. Formally, 
\begin{equation}
{\rm PDF}\,(t\,|\,c,\sigma_c) = \frac{P\,(t)\,P\,(c\,|\,t,\sigma_c)}{P\,({\rm s})\,P\,(c\,|\,{\rm s}, \sigma_c) + P\,({\rm g})\,P\,(c\,|\,{\rm g}, \sigma_c)},\label{pdf_pt}
\end{equation}
where $P\,(t)$ is the prior information for type $t$, and $P\,(c\,|\,t, \sigma_c)$ is the probability of getting the observed concentration under the hypothesis that the source is of type $t$. The PDF is normalised to one by definition,
\begin{equation}
\sum_t {\rm PDF}\,(t) = {\rm PDF}\,({\rm s}) + {\rm PDF}\,({\rm g}) = 1.
\end{equation}
\citet{henrion11} present all the details about the general estimation of the terms in Eq.~(\ref{pdf_pt}). In our case, we estimated the probability of the observations as
\begin{equation}
P\,(c\,|\,t,\sigma_c) = \int \! D_t\,(c_0)\,P_G\,(c\,|\,c_0, \sigma_c)\,{\rm d}c_0,\label{eq_pdt}
\end{equation}
where $D_t$ is the intrinsic distribution (i.e. unaffected by observational errors) of type $t$ sources, and $c_0$ is the real value of the concentration affected by a Gaussian uncertainty
\begin{equation}
P_G\,(x\,|\,x_0, \sigma_x) = \frac{1}{\sqrt{2 \pi} \sigma_x} \exp\bigg[-\frac{(x - x_0)^2}{2\sigma_x^2}\bigg].
\end{equation}
We have no access to the real value of the concentration $c_0$, so we marginalize over it in Eq.~(\ref{eq_pdt}) and the probability is expressed therefore with measured quantities.

The uncertainty in the concentration parameter is
\begin{equation}
\sigma_c = \sqrt{\sigma^2_{r_{1.5^{\prime\prime}}} + \sigma^2_{r_{3.0^{\prime\prime}}} - 2\,\rho\,\sigma_{r_{1.5^{\prime\prime}}}\sigma_{r_{3.0^{\prime\prime}}}},\label{cerr}
\end{equation}
where $\rho$ is the covariance between both magnitudes. We note that the flux of the $1.5^{\prime\prime}$ aperture is included in the $3.0^{\prime\prime}$ aperture, so both measurements are correlated. We have estimated the best covariance value empirically by studying the properties of the stellar locus at different magnitudes (Sect.~\ref{slocus}), and find $\rho = 0.72$. Hereafter, we have assumed such covariance value in our analysis.

Some simplifications were made in the estimation of the ${\rm PDF}$. We neglected the uncertainty in the reference $r-$band \texttt{AUTO} magnitude, that also correlates with the measured concentration as both share common flux. With this simplification, we downgraded the dimensionality of the analysis, passing from a joint 2D estimation in $r - c$ space to a 1D estimation in $c$, and speed up the numerical efficiency of the process. In addition, we did not use the multi-filter colour information from the twelve J-PLUS bands. Such information is highly valuable, and will be included in future versions of the classifier. Despite the above simplifications, our results demonstrate that the implemented version of the classifier has reached a fair compromise between mathematical rigour and computational resources.


The next sections are devoted to the estimation, using J-PLUS EDR data, of the intrinsic distributions for compact ($D_{\rm s}$, Sect.~\ref{slocus}) and extended sources ($D_{\rm g}$, Sect.~\ref{glocus}). The prior probability $P\,(t)$ is estimated in Sect.~\ref{prior}. The final PDF is defined as the combination of the morphological information in the $g$, $r$, and $i$ broad bands, as detailed in Sect.~\ref{pgri}.

\begin{figure*}[t]
\centering
\resizebox{0.49\hsize}{!}{\includegraphics{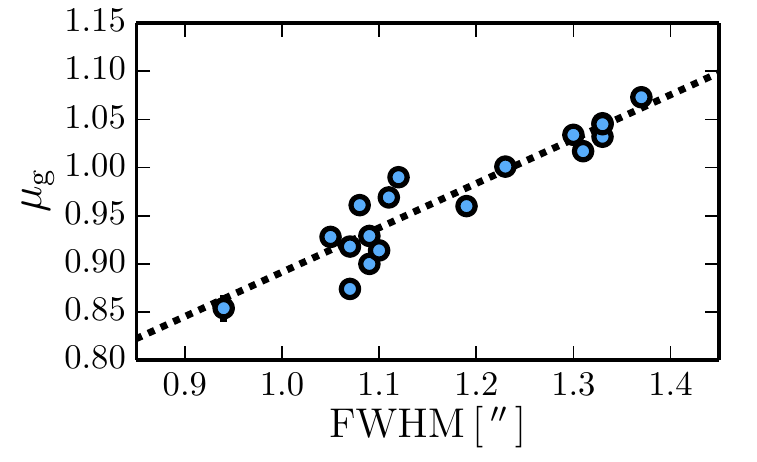}}
\resizebox{0.49\hsize}{!}{\includegraphics{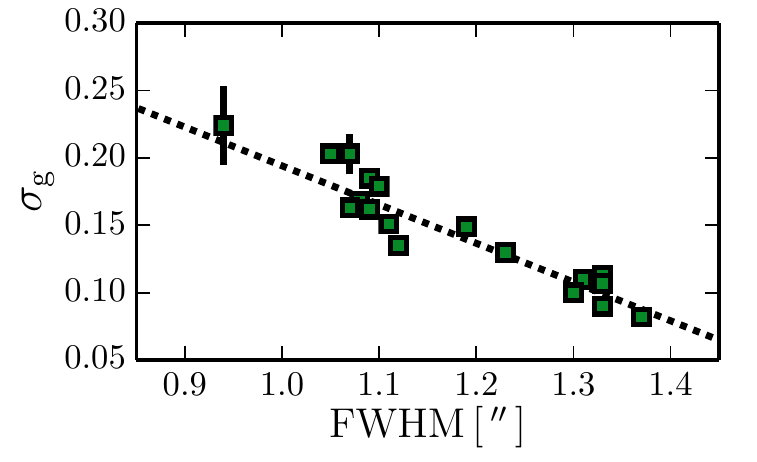}}
\caption{Variation of the parameters defining the galaxy locus (position $\mu_{\rm g}$ in {\it left panel} and dispersion $\sigma_{\rm g}$ in {\it right panel}) as a function of the PSF FWHM of the J-PLUS EDR $r-$band image. The dotted line in each panel is the error-weighted linear fit to the points.}
\label{glocus_fig}
\end{figure*}

\subsection{Definition of the stellar locus}\label{slocus}
The first step of our classifier is to estimate the intrinsic distribution of compact sources in concentration space, noted as stellar locus and defined by $D_{\rm s}$. As shown in Fig.~\ref{magcon_sdss}, the compact population presents a clear, narrow distribution with an extended tail towards larger values of $c_{r}$. This extended component is present at all magnitudes and it is a common feature in astronomical images \citep[see][for details]{henrion11}. To account for this fact, we parametrised the stellar locus as
\begin{equation}
D_{\rm s}\,(c_0\,|\,\boldsymbol{\theta}_{\rm s}) = P_G\,(c_0\,|\,\mu_{\rm s}, \sigma_{\rm s})\,\bigg[1 + {\rm erf}\,\bigg(\alpha_{\rm s}\,\frac{c_0 - \mu_{\rm s}}{\sqrt{2} \sigma_{\rm s}}\bigg)\bigg],\label{dslocus}
\end{equation}
where $\boldsymbol{\theta}_{\rm s} = (\mu_{\rm s}, \sigma_{\rm s}, \alpha_{\rm s})$ are the three parameters that define the distribution, and ${\rm erf}\,(x)$ is the error function. The parameter $\alpha_{\rm s}$ determines the skewness of the distribution, which accounts for the extended tail at larger concentration values.

The stellar locus distribution is affected by the uncertainties in the measured concentrations. To derive the parameters $\boldsymbol{\theta}_{\rm s}$, we maximized the likelihood
\begin{equation}
{\mathcal L}_{\rm s}\,(\vec{c}\,|\,\boldsymbol{\theta}_{\rm s}, \boldsymbol{\sigma}_{c}) = \prod_k \int \! D_{\rm s}\,(c_0\,|\,\boldsymbol{\theta}_{\rm s})\,P_G\,(c_k\,|\,c_0, \sigma_{c,k})\,{\rm d}c_0,\label{convslocus}
\end{equation}
where $\vec{c} = (c_1,c_2,\dotsc,c_k)$ is the data vector with associated uncertainty vector $\boldsymbol{\sigma}_{c}$, and the index $k$ spans the sources in the sample. We used J-PLUS sources with $15 < r \leq 18$ to model the stellar locus because at these magnitudes photometric errors are small and the extended population is sparse. We performed an initial Gaussian fit to the histogram of these sources, and discarded those with large $c_r$ values beyond $5\sigma$ to avoid the few extended sources that exist at these magnitudes. The stellar locus distribution in Eq.~(\ref{dslocus}) has three parameters, but we found that the value of $\alpha_{\rm s}$ is similar in all the pointings. Hence, we fixed it to the J-PLUS EDR median value, $\alpha_{\rm s} = 4.1$, and estimated the values of $\mu_{\rm s}$ and $\sigma_{\rm s}$ in each pointing independently.

We present the dependence of the stellar locus position ($\mu_{\rm s}$) and dispersion ($\sigma_{\rm s}$) with the PSF FWHM of the J-PLUS EDR $r-$band images in Fig.~\ref{slocus_fig}. We found correlations with the FWHM, as expected. The position of the stellar locus increases with the FWHM, implying a less concentrated distribution of light. However, this correlation is affected by a significant scatter, meaning that other factors (e.g. the FWHM of the individual combined images and the stability of the PSF across the large T80Cam FoV) are also modifying $\mu_{\rm s}$ from pointing to pointing.

We found that the value of $\sigma_{\rm s}$ seems to decrease with the PSF FWHM. This trend is caused by the different impact of the PSF depending on the compactness of the source. The concentration of an extended source is less affected by the PSF, so the larger the FWHM, the smaller the difference between compact and extended sources and the smaller the dispersion of the distribution (i.e. the PSF tends to homogenise the concentration of the imaged sources). We confirm this interpretation with the study of the galaxy locus in Sect.~\ref{glocus}. We also note that three pointings are clearly above the general trend, with $\sigma_{\rm s} \sim 0.07$ (Fig.~\ref{slocus_fig}). These pointings were observed at the same night and present a larger PSF FWHM dispersion across the FoV ($\sigma_{\rm FWHM} \sim 0.06$) than the other 15 fields ($\sigma_{\rm FWHM} \sim 0.03$). This highlights the importance of a pointing-by-pointing analysis and the flexibility of our modelling.

We conclude this section by testing the impact of the observational errors in our analysis. The intrinsic (i.e. unaffected by uncertainties) stellar locus must be independent of the magnitude for a given pointing. However, the observed stellar locus broadens towards fainter magnitudes because our concentration measurements have larger uncertainties. We illustrate this fact in Fig.~\ref{slocus_mag} by using the estimated position and dispersion of the stellar locus at different magnitudes, normalized to the fiducial value estimated at $15 < r \leq 18$ for comparison. We found that the stellar locus position and dispersion are consistent in the whole magnitude range only if the concentration uncertainty is included in the analysis via Eq.~(\ref{convslocus}). Disregarding the effect of the errors produces a bias in the position ($\sim 2$\% underestimation at $r \sim 19$) and the dispersion ($\sim 60$\% overestimation at $r \sim 19$) of the stellar locus. We also find that leaving out the covariance term in the estimation of the concentration error (Eq.~[\ref{cerr}]) also biases the results by producing a much lower $\sigma_{\rm s}$ than expected because of the overestimated uncertainties. In practice, we used this fact to estimate empirically the best value for the covariance term in Eq.~(\ref{cerr}), $\rho = 0.72$, so it is an extra parameter in our modelling. We conclude that a rigorous treatment of the concentration uncertainties is fundamental to perform a robust analysis and obtain a meaningful PDF. 

With the stellar locus properly defined, the next step is to repeat the above procedure to estimate the intrinsic distribution of extended sources.

\begin{figure}[t]
\centering
\resizebox{\hsize}{!}{\includegraphics{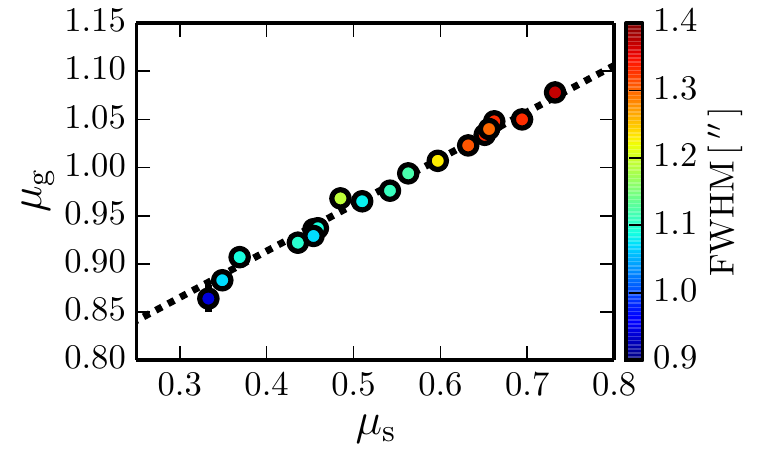}}
\caption{Position of the galaxy locus, $\mu_{\rm g}$, as a function of the position of the stellar locus, $\mu_{\rm s}$. The colour scale shows the PSF FWHM of the $r-$band images. The dotted line is the error-weighted linear fit to the points.}
\label{sglocus}
\end{figure}

\subsection{Definition of the galaxy locus}\label{glocus}
The second step of our classifier is to estimate the intrinsic distribution of extended sources in concentration space, noted as galaxy locus and defined by $D_{\rm g}$. We return to Fig.~\ref{magcon_sdss} to define the shape of the galaxy locus. As \citet{henrion11}, we found that a log-normal distribution is a fair representation of the observations,
\begin{equation}
D_{\rm g}\,(c_0\,|\,\boldsymbol{\theta}_{\rm g}) = \frac{1}{c_0 \sqrt{2 \pi} \sigma_{\rm g}} \exp\bigg[-\frac{(\ln c_0 - \ln \mu_{\rm g})^2}{2\sigma_{\rm g}^2}\bigg],
\end{equation}
where $\boldsymbol{\theta}_{\rm g} = (\mu_{\rm g}, \sigma_{\rm g})$ are the two parameters that define the distribution. We assume that these parameters do not depend on source magnitude. This seems a good approximation in our magnitude range, $r < 21$, but it is not guaranteed at fainter magnitudes. Thus, a magnitude-dependent parametrisation of the galaxy locus could be needed for deeper surveys.

We derived the parameters $\boldsymbol{\theta}_{\rm g}$ by maximizing the likelihood
\begin{align}
{\mathcal L}_{\rm g}\,&(\vec{c}\,|\,\boldsymbol{\theta}_{\rm g}, f_{\rm g}, \boldsymbol{\sigma}_{c}) = \nonumber\\
&\prod_k \!\! \int \!\! \big[(1 - f_{\rm g})\,D_{\rm s}\,(c_0) + f_{\rm g}\,D_{\rm g}\,(c_0\,|\,\boldsymbol{\theta}_{\rm g})\big]\,P_G\,(c_k\,|\,c_0, \sigma_{c,k})\,{\rm d}c_0,\label{lglocus}
\end{align}
where $f_{\rm g}$ is the fraction of galaxies in the sample, and the parameters of the stellar locus were fixed to the values derived in Sect~\ref{slocus}. In the modelling of the galaxy locus, we used sources with $18 < r \leq 20$, where both compact and extended sources exists. This magnitude range ensures well controlled uncertainties, a significant separation between extended and compact populations, and enough number of extended sources to derive the parameters, with inferred values of $f_{\rm g}$ between 0.20 and 0.55 at $18 < r \leq 20$. We show the location of the galaxy locus and its dispersion as a function of the pointing PSF FWHM in Fig.~\ref{glocus_fig}. As in the case of the stellar locus parameters, $\mu_{\rm g}$ increases with FWHM and $\sigma_{\rm g}$ decreases.

We compare the location of the stellar and the galaxy loci in Fig.~\ref{sglocus}. We found that both variables are highly correlated with a slope below unity, ${\rm d}\mu_{\rm g}/{\rm d}\mu_{s} = 0.48$. This implies that the increase of the PSF FWHM affects more the compact sources, and for a large enough FWHM both populations would be indistinguishable in our images. This should occur with a FWHM larger than $\sim2.5^{\prime\prime}$. This effect also explains the decrease of $\sigma_{\rm s}$ and $\sigma_{\rm g}$ with FWHM, as argued in Sect.~\ref{slocus}.

We conclude that our modelling with six parameters (three for the stellar locus, two for the galaxy locus, and the magnitude covariance) is general enough to retain the main properties of compact and extended sources. A pointing-by-pointing analysis is needed to account for the different observational conditions that modify the position and the dispersion of the stellar and galaxy loci. The PSF FWHM of the images is the main parameter involved, but it cannot explain all the observed variations alone. In the next section, we compute the final ingredient of our classifier, the prior in the fraction of stars and galaxies with magnitude.

\begin{figure}[t]
\centering
\resizebox{\hsize}{!}{\includegraphics{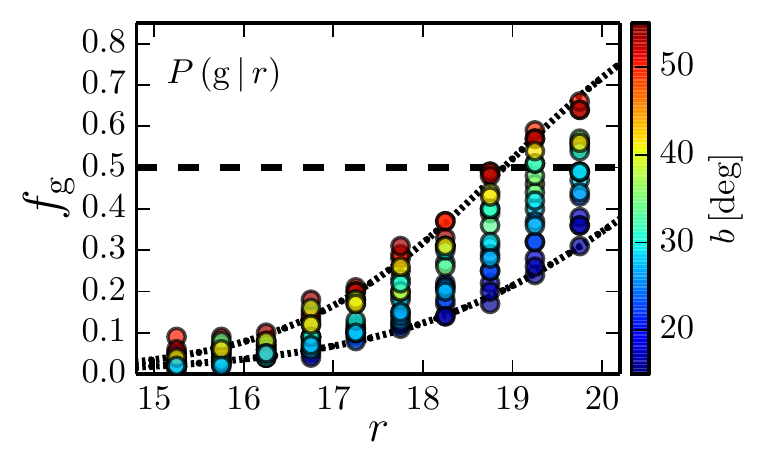}}
\caption{Fraction of galaxies, $f_{\rm g}$, as a function of the $r-$band magnitude. The circles are the measurements in each J-PLUS EDR pointing, coloured with the galactic latitude $b$ of the pointing. The dashed line marks $f_{\rm g} = 0.5$. As illustration, the dotted lines show the best prior fitting curves, $P\,({\rm g}\,|\,r)$, to the pointings with lower and larger galactic latitude.   
}
\label{fgal_edr}
\end{figure}

\begin{figure}[t]
\centering
\resizebox{\hsize}{!}{\includegraphics{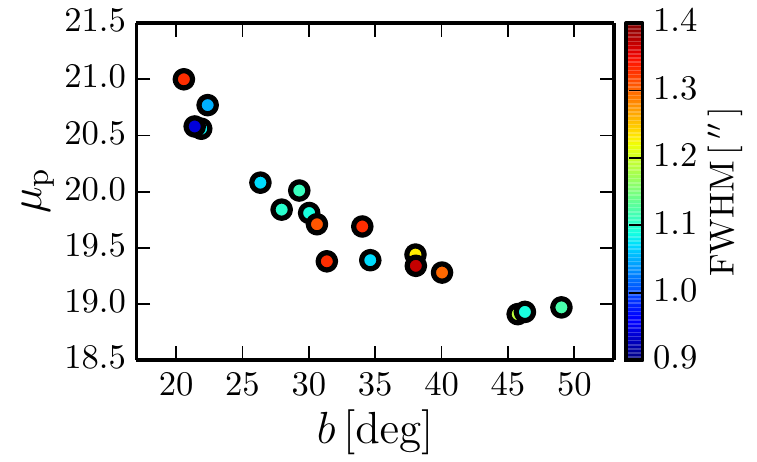}}
\caption{Magnitude at which stars and galaxies have the same number density, $\mu_{\rm p}$, as a function of the galactic latitude $b$ of the J-PLUS EDR pointings. The colour scale shows the FWHM of the $r-$band images.}
\label{mup_bgal}
\end{figure}

\begin{figure*}[t]
\centering
\resizebox{0.495\hsize}{!}{\includegraphics{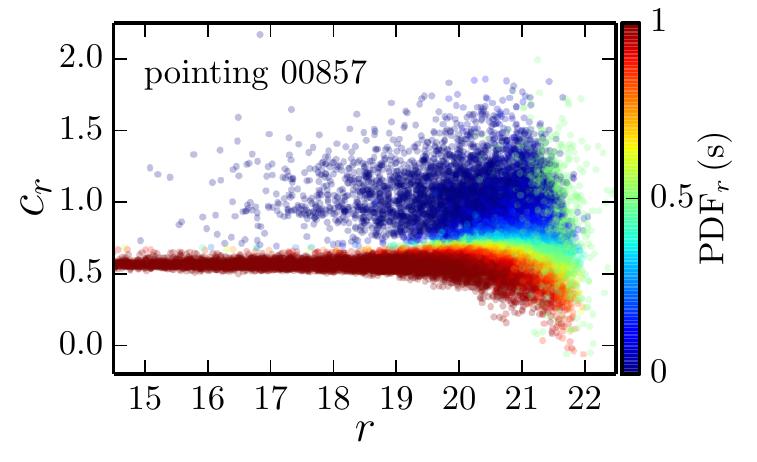}}
\resizebox{0.495\hsize}{!}{\includegraphics{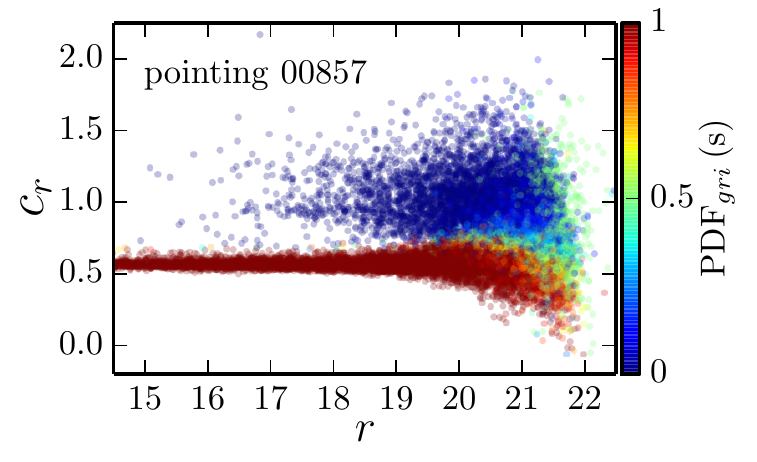}}
\caption{Stellar probability in the magnitude vs. concentration plane for the J-PLUS EDR pointing 00857. The {\it left panel} shows the classification derived with the $r-$band image alone, ${\rm PDF}_{r}\,({\rm s})$, and the {\it right panel} the final classification combining the morphological information from $g$, $r$, and $i$ bands, ${\rm PDF}_{gri}\,({\rm s})$.}
\label{magcon_ps}
\end{figure*}

\subsection{Prior probability by morphological type}\label{prior}
We have defined the properties of the stellar and the galaxy loci in the previous sections at magnitudes brighter than $r = 20$, where both populations can be distinguished. We aim to define a probabilistic classification that can also be useful and statistically meaningful at magnitudes close to our detection limit, $r \sim 21$. To reach our goal, we have to include a prior information about the relative number of stars and galaxies with magnitude \citep{scranton02,henrion11,molino13}. This is the third step of our classifier.

We computed the fraction of galaxies at different magnitudes, from $r = 15.25$ to $r = 19.75$ in 0.5 mag steps, by minimising Eq.~(\ref{lglocus}) with only $f_{\rm g}$ as free parameter (i.e. we fixed the stellar and galaxy loci parameters). The derived galaxy fractions for each J-PLUS EDR pointing are shown in Fig.~\ref{fgal_edr}. We found a variety of values, with pointings closer to the Milky Way disk having a lower fraction of galaxies. To have a continuum description with magnitude and prior information beyond $r = 20$, we fitted the dependence of $f_{\rm g}$ on $r-$band magnitude as
\begin{equation}
f_{\rm g}\,(r\,|\,\boldsymbol{\theta_{\rm p}}) = \frac{1}{1 + {\rm exp}[ -\kappa_{\rm p}\,(r - \mu_{\rm p}) ] },
\end{equation}
where $\boldsymbol{\theta_{\rm p}} = (\mu_{\rm p}, \kappa_{\rm p})$ are the parameters of the function, $\mu_{\rm p}$ is the magnitude with the same number density of stars and galaxies, and $\kappa_{\rm p}$ is the transition rate between both populations. 

We defined the prior probability of a source with magnitude $r$ as $P\,({\rm g}\,|\,r) = f_{\rm g}\,(r)$ for galaxies, and as $P\,({\rm s}\,|\,r) = 1 - f_{\rm g}\,(r)$ for stars. The prior information ensures statistical meaningful probabilities at magnitudes fainter than $r = 20$, where less information is encoded in the concentration parameter because of observational uncertainties.

We show the derived values of $\mu_{\rm p}$ as a function of the galactic latitude $b$ of the J-PLUS EDR pointings in Fig.~\ref{mup_bgal}. We find that $\mu_{\rm p}$ increases as we approach the galactic plane, as expected. The parameter $\mu_{\rm p}$ has no clear dependence on the PSF FWHM, supporting our modelling procedure. We note that with the pointing-by-pointing estimation of $P\,(t)$, we had implicitly included in our prior probability the varying density of stars with the position in the sky \citep{scranton02}.

Finally, to avoid biases related to low-quality and undetected sources, we only assigned a probability to those detections with ${\rm S/N} > 3$. Below this ${\rm S/N}$ limit, measured fluxes are effected by the sky background fluctuations and the concentration measurements are compromised. The low signal-to-noise sources with ${\rm S/N} \leq 3$ have ${\rm PDF}\,({\rm s}) = 0.5$, irrespective of their likelihoods and magnitude.

\subsection{Final probability from $g$, $r$, and $i$ bands information}\label{pgri}
In addition to the concentration in the $r$ band, we included in our final classification the valuable information from the $g-$ and $i-$band images. These three broad bands are the deeper ones in our dataset, and provide most of the morphological J-PLUS information.

The stellar and the galaxy loci in $g$ and $i$ bands were estimated following Sects.~\ref{slocus} and \ref{glocus}, adapting the analysed magnitude ranges to the depth and properties of each band. 

We assumed uninformative flat priors for $g$ and $i$ bands, $P\,(t\,|\,g) = P\,(t\,|\,i) = 0.5$, in Eq.~(\ref{pdf_pt}). Stars and galaxies have particular colour distributions with $r-$band magnitude, as we show in Sect.~\ref{pdfgi}. Thus, $g$, $r$, and $i$ magnitudes are not independent and a complete multi-filter study demands the inclusion of the colour distributions for different morphological types in the probability from Eq.~(\ref{eq_pdt}), as shown by \citet{henrion11}. This will include explicitly the colour information in the analysis, but we only considered here the morphological one by imposing flat priors and using no colours. The addition of the colour information from the twelve J-PLUS bands is beyond the scope of the present paper, and will be included in future versions of the classifier.

Regarding the quality cut at ${\rm S/N} = 3$, it was applied in each band independently. This allows a meaningful inclusion of uninformative bands, and only the relevant information was used to perform the final classification of the sources.

We estimated the final PDFs by multiplying the probabilities from each band and normalising to one. We illustrate the impact of the $gri$-derived classification with respect to the $r-$band one in Fig.~\ref{magcon_ps}. We show the star probability in the magnitude vs. concentration plane for the representative pointing 00857. The other pointings in the J-PLUS EDR present a similar behaviour, and all their relevant figures are accessible at the PROFUSE web page. In the single-band case, the classification performs well up to $r \sim 19.5$, the magnitude at which the uncertainties in the concentration parameter start to merge the two populations. At fainter magnitudes both distributions overlap, creating a continuous transition between stars and galaxies. This performance is common among Bayesian classifiers \citep{henrion11}. The inclusion of the extra morphological information from the $g$ and $i$ bands is evident in the {\it right panel} of Fig.~\ref{magcon_ps}, where the continuous transition in the overlapping area is diminished because those sources that appear compact in the $g$ and $i$ bands approach to ${\rm PDF}\,({\rm s}) = 1$. We further study the concentration distribution of stars and galaxies in Sect.~\ref{pdfc}.

At this point, we have a PDF-based probabilistic classification of J-PLUS EDR sources into star or galaxy. In the next sections, we test our classification by comparison with SDSS dataset in the common areas.

\begin{figure*}[t]
\centering
\resizebox{0.49\hsize}{!}{\includegraphics{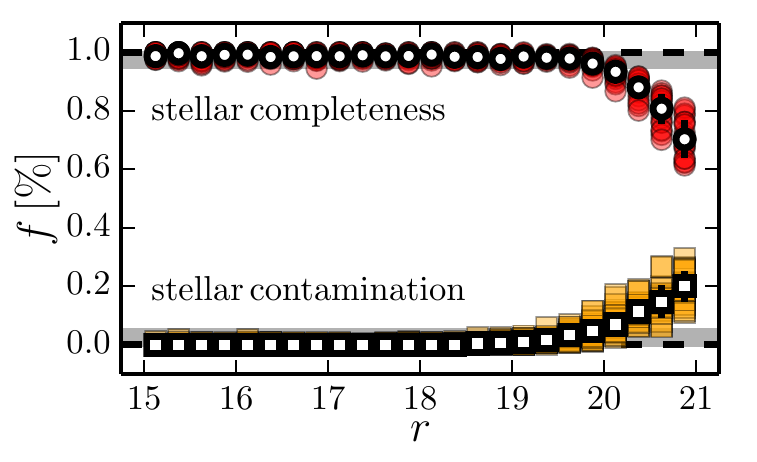}}
\resizebox{0.49\hsize}{!}{\includegraphics{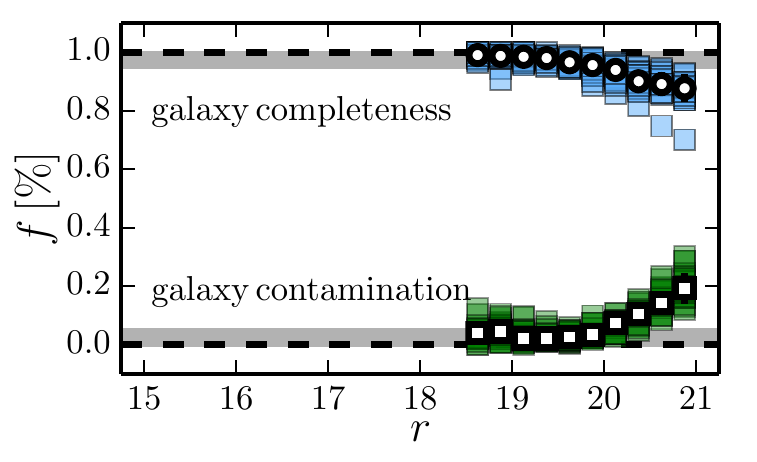}}
\caption{Completeness (circles) and contamination (squares) of the J-PLUS EDR stars ({\it left panel}) and galaxies ({\it right panel}) as a function of the $r-$band magnitude. Sources were classified using ${\rm PDF}_{gri} = 0.5$ as boundary. The colour symbols show the estimation on individual J-PLUS EDR pointings. The white symbols are the median values of the pointings. The grey areas mark the 95\% completeness and 5\% contamination tolerance regions used to define the performance of our classifier.}
\label{compu_edr}
\end{figure*}

\section{Stars and galaxies in the J-PLUS EDR}\label{test}
We tested the performance of our PDF-based morphological classifier implemented in Sect.~\ref{psjplus} by comparing it to SDSS dataset. The SDSS classification is reliable up to $r \sim 21$ \citep{yasuda01,scranton02}, and it is used as reference in the next sections. We stress that our classification was based solely on J-PLUS information and it is independent of SDSS classification.

As a first step, we performed a one-to-one comparison with the SDSS classification, defining the completeness and the contamination of our classifier (Sect~\ref{compu}). We stress the probabilistic nature of our PDF-based classification by studying the distribution of stars and galaxies in both concentration (Sect.~\ref{pdfc}) and $(g-i)$ colour space (Sect.~\ref{pdfgi}) at different magnitudes.

\subsection{Completeness and contamination for a boolean classification}\label{compu}
The usual way to test the performance of a classifier is analysing the completeness and contamination of its output classes. The completeness is defined as the fraction of sources classified with type $t$ over the total number of actual sources of type $t$. The contamination is the fraction of sources wrongly classified as type $t$.

We present the completeness and contamination of our Bayesian classifier in Fig.~\ref{compu_edr}. To construct this figure, we used a boolean classification with stars as ${\rm PDF}\,({\rm s}) > 0.5$ sources. We found that this boolean classification is reliable up to $r \sim 20$, with a completeness of $\sim 95$\% and a contamination of $\sim5$\% both for stars and galaxies. Thus, for those studies that need a secure source type $t$, J-PLUS data provide a reliable morphological classification at $r \lesssim 20$. However, our PDF-based classifier is suited for statistical studies, and we demonstrate its good performance up to $r = 21$ in the following sections.

\subsection{Distribution of stars and galaxies in concentration space}\label{pdfc}
\begin{figure*}[t]
\centering
\resizebox{0.49\hsize}{!}{\includegraphics{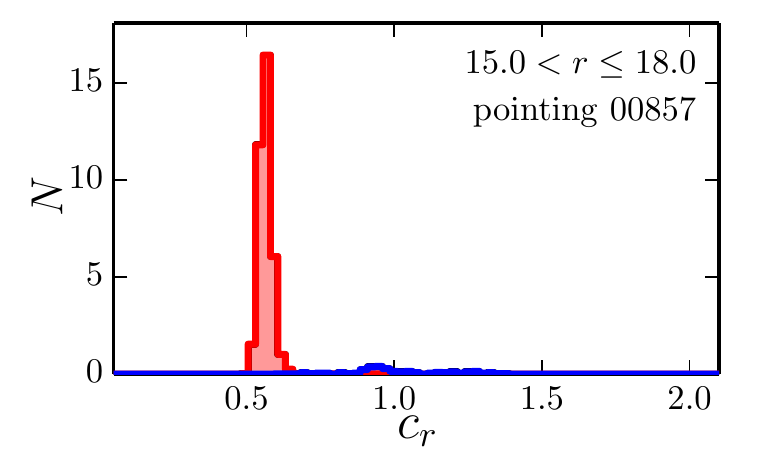}}
\resizebox{0.49\hsize}{!}{\includegraphics{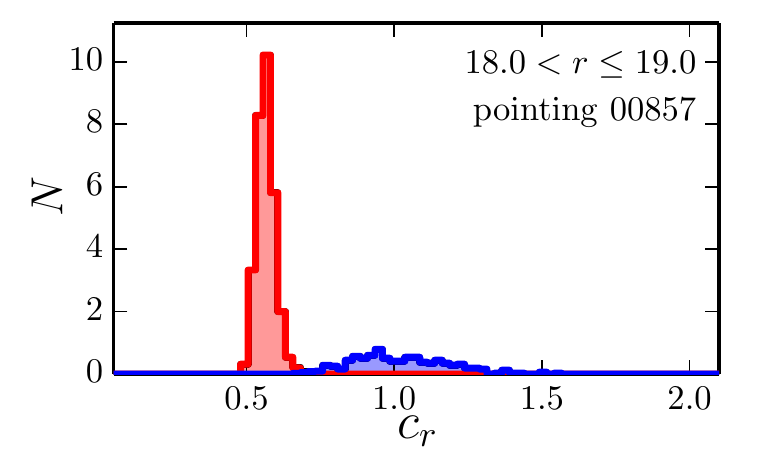}}\\
\resizebox{0.49\hsize}{!}{\includegraphics{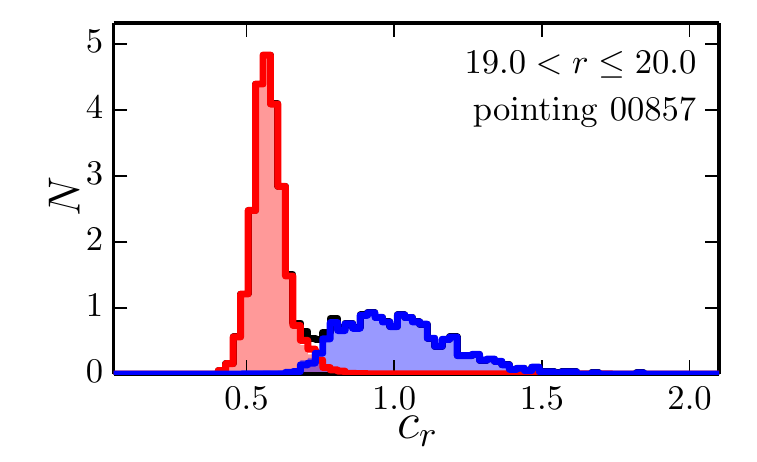}}
\resizebox{0.49\hsize}{!}{\includegraphics{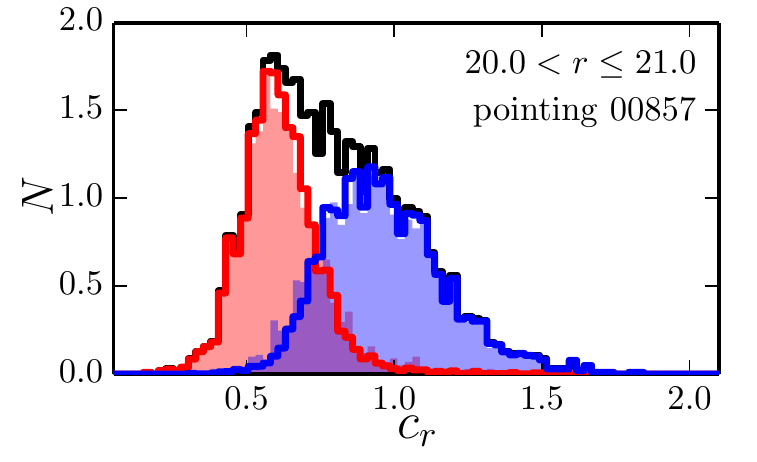}}
\caption{Distribution of stars (red) and galaxies (blue) in the $r-$band concentration space for different magnitude ranges (labelled in the panels) in the J-PLUS EDR pointing 00857. The coloured solid lines mark the PDF-weighted histograms estimated with our probabilistic, PDF-based classifier. The coloured areas show the histograms estimated with the SDSS classification of J-PLUS sources. The black solid histogram is the total distribution, illustrating the confusion between compact and extended sources at the faintest magnitudes.}
\label{con_ps}
\end{figure*}

We present the PDF-weighted distribution of star and galaxies in pointing 00857 for different magnitude ranges in Fig.~\ref{con_ps}. We also present the distribution obtained with the SDSS classification of our sources, similar to the {\it bottom panel} in Fig.~\ref{magcon_sdss}. We found that the SDSS distribution is well recovered by our probabilistic approach. This agreement is remarkable at the fainter magnitude range with $20 < r \leq 21$. The stellar locus in concentration space is broad at these magnitudes and stellar sources extend to larger than expected concentrations because of the uncertainties in $c_r$. This fact is confirmed by the SDSS-based classification, and correctly accounted for with our statistical analysis.

\begin{figure*}[t]
\centering
\resizebox{0.49\hsize}{!}{\includegraphics{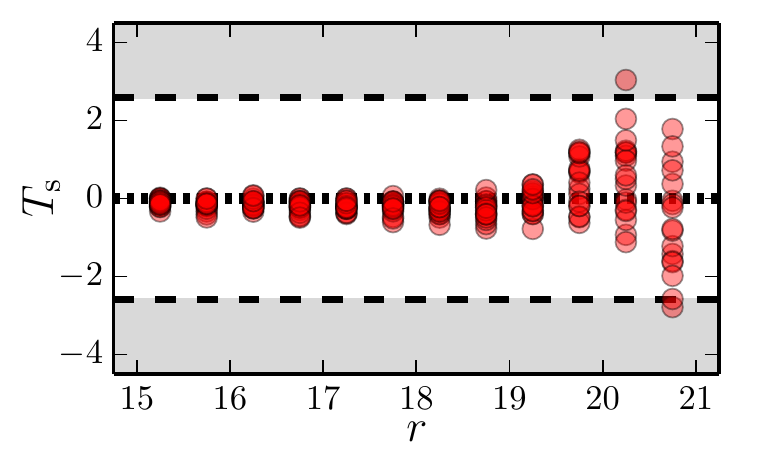}}
\resizebox{0.49\hsize}{!}{\includegraphics{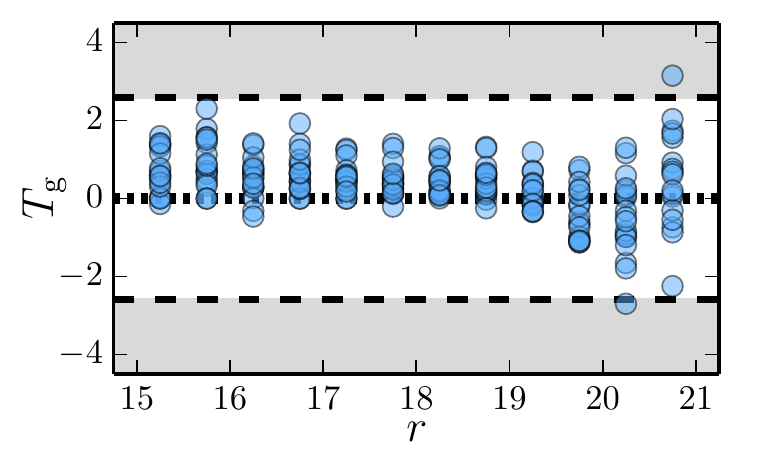}}
\caption{Significance of the difference between the number of PDF-based stars ({\it left panel}) and galaxies ({\it right panel}) with respect to the number estimated from SDSS classification, $T_{t}$, as a function of $r-$band magnitude. The coloured symbols show the estimation on individual J-PLUS EDR pointings. The dotted lines mark zero difference. The dashed lines mark $|T| = 2.6$, the limit to consider the PDF-based and the SDSS-based numbers as similar. The exclusion areas with $|T| > 2.6$ are shown in grey.}  
\label{R_edr}
\end{figure*}

The previous qualitative comparison based on pointing 00857 is quantitatively explored in Fig.~\ref{R_edr}. We compared at different magnitudes ranges the number of SDSS-based stars and galaxies, $N_{\rm SDSS}^t$, with the number estimated from the addition of the morphological type PDFs, 
\begin{equation}
N_{\rm PDF}^t = \sum_k {\rm PDF}_k\,(t).
\end{equation}
We carried out a Student's t-test to state the significance of the difference between both estimations, assuming $N_{\rm SDSS}^t$ as the right one. The estimator is defined as
\begin{equation}
T_{t} = \frac{N_{\rm PDF}^t - N_{\rm SDSS}^t}{\sqrt{N_{\rm PDF}^t}},
\end{equation}
where Poissonian uncertainties were assumed. We accepted that $N_{\rm PDF}^t = N_{\rm SDSS}^t$ with $\alpha = 0.01$ significance at $|T_t| \leq 2.6$ (i.e. 99\% of the time two similar distributions randomly sampled fulfils this criteria). We found that nearly all the measurements are compatible, with only two discrepant pointings at the faint end. Thus, we conclude that the PDF-based classifier statistically obtains the right number of stars and galaxies up to $r = 21$, in contrast with the limit $r = 20$ imposed by a boolean classification (Sect.~\ref{compu}). Hence, by relying on the Bayesian classification and weighting sources with the morphological type PDF instead of using the boolean one, we can extend the analysis to one magnitude deeper. To highlight this fact, we study the $(g-i)$ colour distributions of J-PLUS EDR stars and galaxies in the next section. 

\begin{figure*}[t]
\centering
\resizebox{0.49\hsize}{!}{\includegraphics{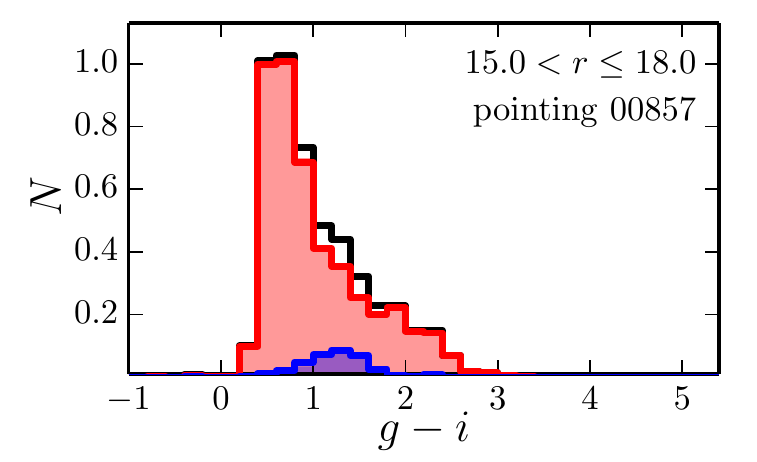}}
\resizebox{0.49\hsize}{!}{\includegraphics{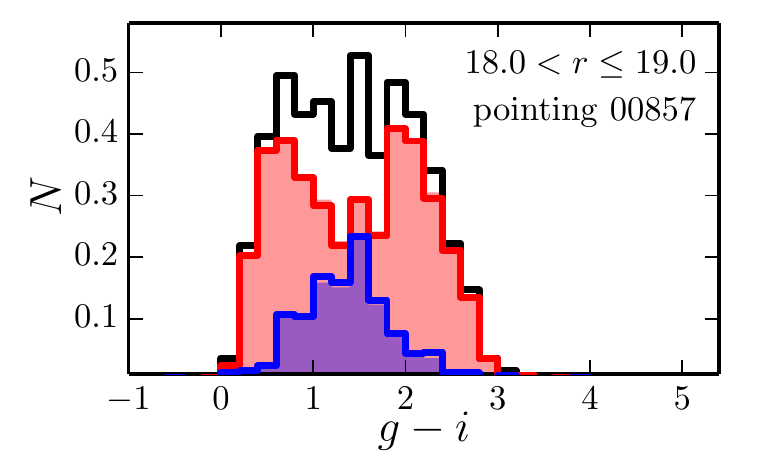}}\\
\resizebox{0.49\hsize}{!}{\includegraphics{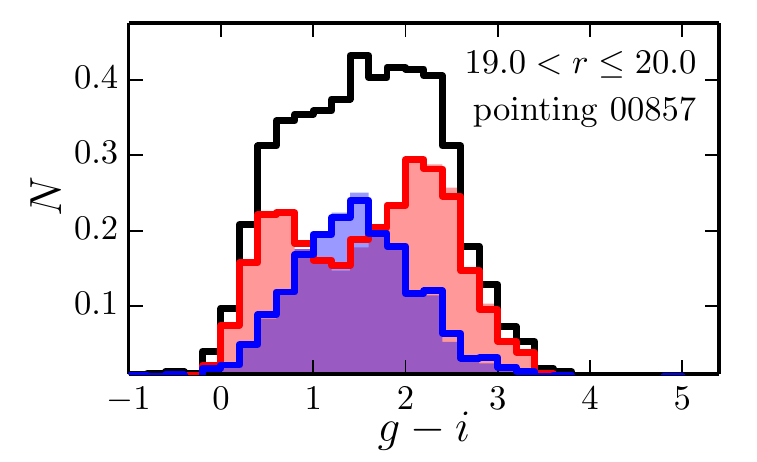}}
\resizebox{0.49\hsize}{!}{\includegraphics{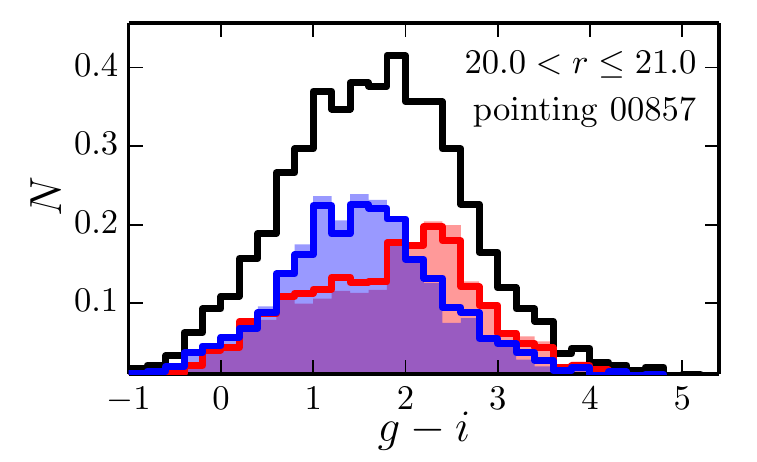}}
\caption{Distribution of stars (red) and galaxies (blue) in the $g-i$ colour space for different $r-$band magnitude ranges (labelled in the panels) at J-PLUS EDR pointing 00857. The solid lines mark the PDF-weighted histograms estimated with our classifier. The coloured areas mark the histograms estimated with the SDSS classification. The black solid histogram is the total colour distribution.}
\label{color_ps}
\end{figure*}

\begin{figure}[t]
\centering
\resizebox{\hsize}{!}{\includegraphics{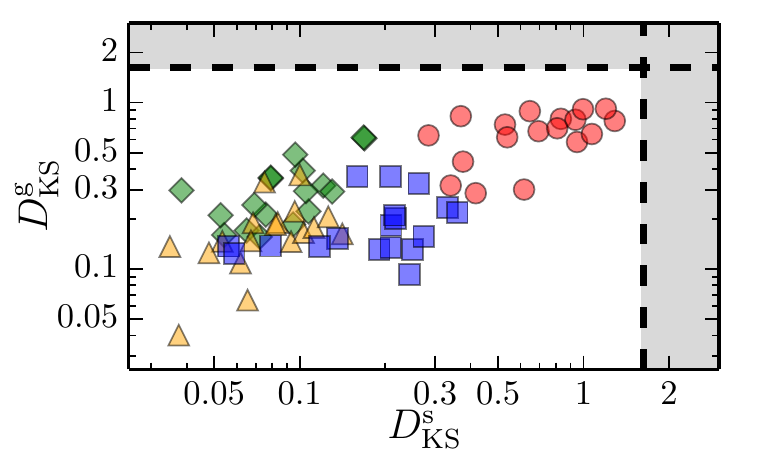}}
\caption{Normalized Kolmogorov-Smirnov parameter for the $(g-i)$ colour distribution of stars ($D_{\rm KS}^{\rm s}$) and galaxies ($D_{\rm KS}^{\rm g}$). The coloured symbols show the estimation on individual J-PLUS EDR pointings. The colours and symbol shapes mark different magnitude ranges, $15 < r \leq 18$ (green diamonds), $18 < r \leq 19$ (yellow triangles), $19 < r \leq 20$ (blue squares), and $20 < r \leq 21$ (red circles). The dashed lines mark $D_{\rm KS} = 1.62$, the limit to consider the PDF-based and the SDSS-based distributions as similar. The exclusion areas with $D_{\rm KS} > 1.62$ are shown in grey.}
\label{Dks_edr}
\end{figure}

\begin{figure*}[t]
\centering
\resizebox{0.49\hsize}{!}{\includegraphics{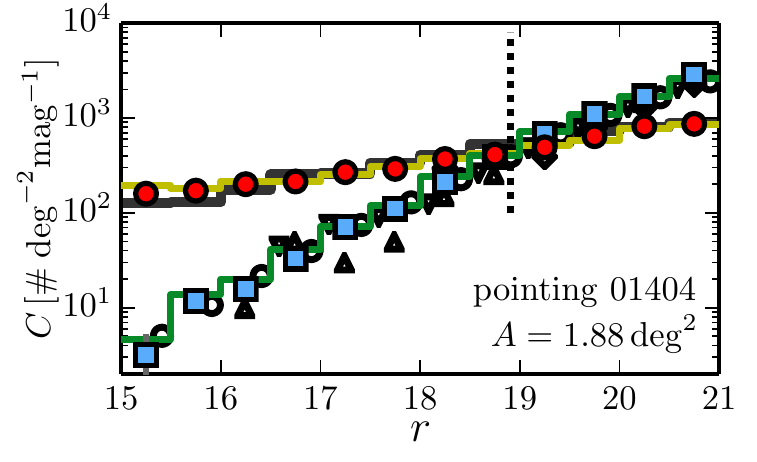}}
\resizebox{0.49\hsize}{!}{\includegraphics{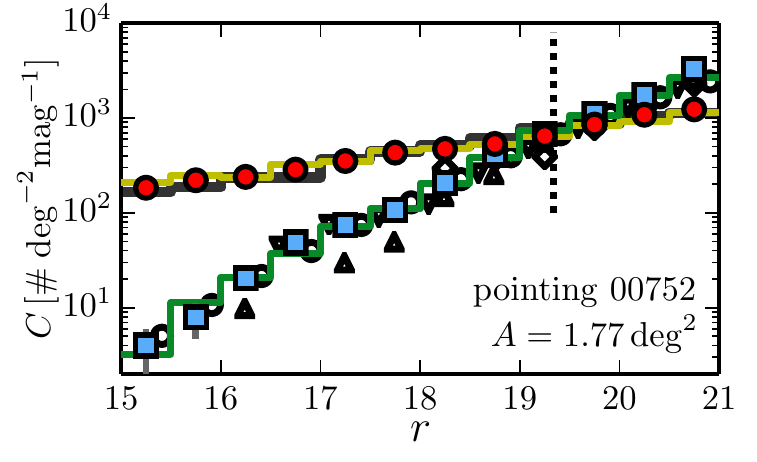}}\\
\resizebox{0.49\hsize}{!}{\includegraphics{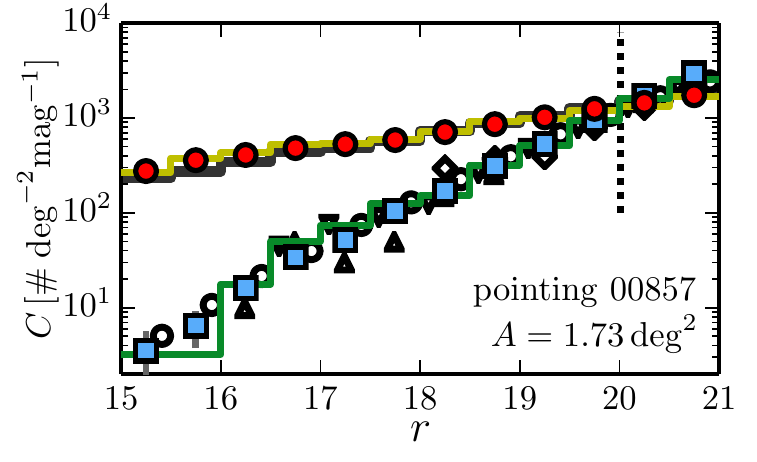}}
\resizebox{0.49\hsize}{!}{\includegraphics{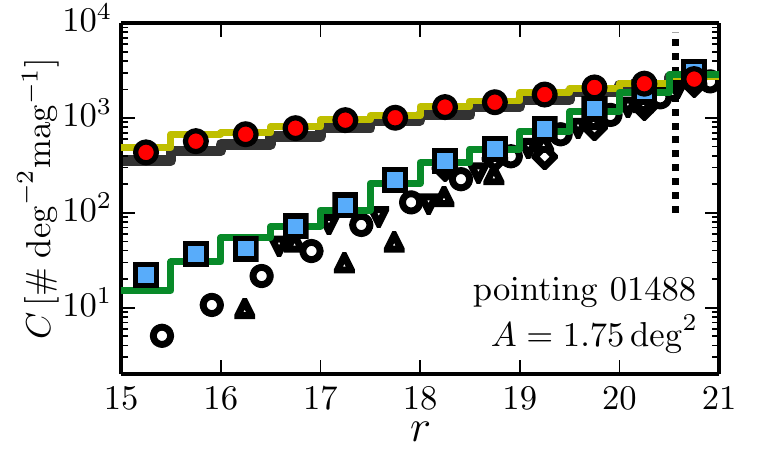}}
\caption{Stellar (red dots) and galaxy (blue squares) number counts as a function of $r-$band magnitude in four J-PLUS EDR pointings (labelled in the panels). Coloured solid histograms are the stellar (yellow) and galaxy (green) number counts from SDSS in the same area. Black solid histograms are the stellar number counts at the pointing position estimated with the TRILEGAL model of the Milky Way \citep{trilegal}. White symbols are galaxy number counts from the literature: \citet[][circles]{yasuda01};  \citet[][triangles]{huang01}; \citet[][inverted triangles]{kummel01}; and \citet[][diamonds]{kashikawa04}. The dashed vertical line marks the value of $\mu_{\rm p}$ estimated for each pointing, showing the expected magnitude with the same number density of stars and galaxies.}
\label{counts_ps}
\end{figure*}

\begin{figure*}[t]
\centering
\resizebox{0.49\hsize}{!}{\includegraphics{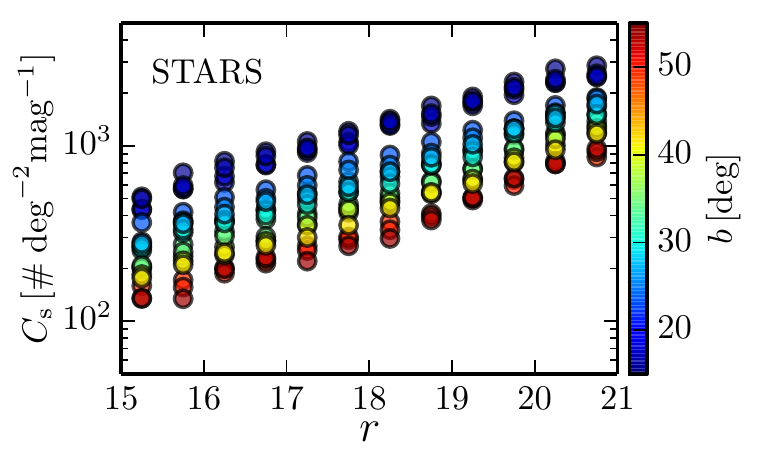}}
\resizebox{0.49\hsize}{!}{\includegraphics{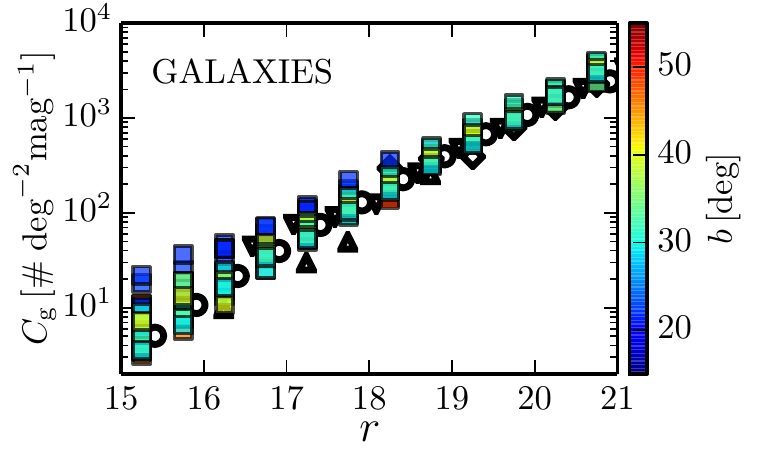}}
\caption{Stellar (dots in the {\it left panel}) and galaxy (squares in the {\it right panel}) number counts as a function of $r-$band magnitude of each J-PLUS EDR pointing. The colour scale shows the galactic latitude of the pointings. White symbols in the {\it right panel} are galaxy number counts from the literature as in Fig.~\ref{counts_ps}.}
\label{counts_edr}
\end{figure*}

\subsection{Colour distribution of stars and galaxies}\label{pdfgi}
We further tested our classifier by computing the distribution of stars and galaxies in ($g-i$) colour space. As in the previous section, we compared our PDF-based distributions with those estimated from the SDSS classification of J-PLUS EDR sources. We present such distributions at different magnitude bins for pointing 00857 in Fig.~\ref{color_ps}. We found that our morphological classification recovers the expected ($g-i$) colour distributions, with a bimodality in the stellar population at all magnitudes. We stress that we are able to recover the expected colour distribution even at $r > 20$, confirming the goodness of our PDF probabilities. We note that the addition of $g$ and $i$ bands to the analysis contributes specially at red colours with $(g-i) > 2$, which are worse recovered with the $r-$band information alone.  

The result of the quantitative comparison of the distributions with a Kolmogorov-Smirnov (KS) test is presented in Fig.~\ref{Dks_edr}. To avoid the different sample sizes, we defined the normalized KS parameter as
\begin{equation}
D_{\rm KS}^t = {\rm max}\big| F^t_{\rm PDF} - F^t_{\rm SDSS} \big| \, \sqrt{\frac{N_{\rm SDSS}^t \ N_{\rm PDF}^t}{N_{\rm SDSS}^t + N_{\rm PDF}^t}},
\end{equation}
where $F^t$ is the cumulative $(g-i)$ colour distribution for type $t$ using a given classification. This parameter is independent of the samples size and provides a common scale for all our measurements. One can consider two distributions to be similar ($\alpha = 0.01$) with $D_{\rm KS} < 1.62$. We found that all the PDF-based colour distributions are compatible with the SDSS-based ones, even at the fainter magnitudes. The inclusion of the $g$ and $i$ bands information improves $D_{\rm KS}$ by a factor of two with respect to only using the $r$ band, confirming the benefits of a multi-filter analysis of the data.

\subsection{Number density of stars and galaxies}\label{stats}
From the results in the previous sections, we conclude that our PDF-based classifier provides a reliable boolean classification at $r \leq 20$, and a meaningful statistical classification at $r \leq 21$. We recover statistically not only the right number of stars and galaxies, but also their colour $(g-i)$ distributions. Any future study with J-PLUS EDR can benefit from our PDF classification by properly weighting the sources in the dataset.

Finally, we estimated the total PDF-weighted number of stars and galaxies with $15 < r \leq 21$ in the high-quality area covered by the J-PLUS EDR (31.70 deg$^2$, Sect.~\ref{mask}). We found 150k stars (4730 per deg$^{2}$) and 101k galaxies (3190 per deg$^{2}$). Assuming these number densities as representative, we expect the detection $\sim40$ million stars and $\sim25$ million galaxies at J-PLUS completion. As a first application, we present the pointing-by-pointing stellar and galaxy number counts in the next section.

\section{Stellar and galaxy number counts}\label{counts}
In this section, we study the number counts in the J-PLUS EDR using our PDF-based classification. We define the $r-$band stellar number counts in pointing $j$ as the PDF-weighted histogram normalised by area and magnitude,
\begin{equation}
C_{\rm s}\,(r_n) = \frac{1}{A_j\,\Delta r} \sum_k \frac{{\rm PDF}_k\,({\rm s})}{f_{\rm c}\,(r_k\,|\,{\rm s})}\ \mathbf{1}\,(r_n - \tfrac{1}{2}\Delta r < r_k \leq r_n + \tfrac{1}{2}\Delta r),
\end{equation}
where $\vec{r} = r_n$ is the $r-$band magnitude vector that defines the histogram, $\Delta r = 0.5$ the magnitude bin size, the index $k$ spans the sources in the pointing, $\mathbf{1}\,(\cdot)$ is the indicator function with value unity if the defined condition is fulfilled and zero otherwise, $A_j$ the area subtended by the pointing, and $f_{\rm c}$ the detection completeness for stars (Sect.~\ref{completeness}). The dimensions of the number counts are [\# deg$^{-2}$ mag$^{-1}$]. We defined the galaxy number counts in the same way, but we weighted with ${\rm PDF}\,({\rm g})$ and used the detection completeness for galaxies. We note that some sources count as stars and galaxies simultaneously in the analysis because of the probabilistic nature of our classification. The uncertainties in the number counts were estimated using the bootstrapping technique \citep{bootstrap}.

We present the J-PLUS EDR stellar and galaxy number counts of four representative pointings in Fig.~\ref{counts_ps}. These pointings span the range of $\mu_{\rm p}$ covered by the data (Sect.~\ref{prior}). The derived stellar and galaxy number counts are in good agreement with SDSS ones up to $r = 21$. Our results are also compatible with the galaxy number counts from the literature and with the stellar number counts from the TRILEGAL models of the Milky Way \citep{trilegal}. The derived number counts can be found at the PROFUSE web page together with their uncertainties.

We further explore the properties of the derived number counts in the following. We focus first on the stellar counts, that are gathered together in the {\it left panel} of Fig.~\ref{counts_edr}. There is a large variation in the normalization of the counts. Using $r  = 19.75$ as reference, the median stellar counts in the J-PLUS EDR is $C_{\rm s}\,(19.75) = 1175 \pm 526$, with $\sim45$\% dispersion. Following the results from Fig.~\ref{mup_bgal}, we found that the number counts decreases with galactic latitude $b$, as we move away from the galactic plane. Moreover, the higher-latitude pointings seems to present a double power-law shape. This can be interpreted as the dominance of halo stars in these fields \citep[e.g.][]{gao13}, but a detailed study is beyond the scope of the present paper.

Regarding the galaxy number counts, the pointing-to-pointing variation is much smaller, with the median counts at $r = 19.75$ being $C_{\rm g}\,(19.75) = 1082 \pm 125$, only a $\sim10$\% dispersion. The pointing-to-pointing scatter is larger at brighter magnitudes, reaching $75$\% at $r = 15.25$. This reflects the larger cosmic variance and points to galaxy clustering as the dominating source of scatter. Indeed, we note that pointings 01488 and 01588 present a clear excess of bright ($r < 17$) sources ({\it bottom right panel} in Fig.~\ref{counts_ps}). We checked that this excess is due to a large scale structure at $z \sim 0.07$, confirmed with SDSS spectroscopy. 

We plan to perform a detailed modelling of the galaxy number counts with the first J-PLUS data release, that will comprise an order of magnitude larger area than the EDR.

\section{Discussion and future prospects}\label{discussion}
From the analysis presented in previous sections, we conclude that meaningful statistical studies of stars and galaxies at $r \leq 21$ with J-PLUS data can be performed by PDF-weighting the observed sources (Sect.~\ref{counts}). Those studies that demand an absolute classification or the spectroscopic follow-up of particular sources, should either restrict to $r \leq 20$ or use high-probability sources with ${\rm PDF}\,(t) > 0.9$. The latter approach permits the identification of reliable candidates up to $r \sim 21$, but penalises the completeness of the targeted morphological population (see \citealt{viironen15}, for an example of this issue regarding the study of high-z galaxies with redshift PDFs).

Our classifier could be improved and further tested by:
\begin{itemize}
\item Including J-PLUS colour information, with twelve available bands from 3500$\AA$ to 9000$\AA$ (Table~\ref{tab:JPLUS_filters}), providing eleven independent colours with valuable information to classify sources;

\item Calibrating with deeper and well defined star/galaxy classifications. Our derived PDFs can be further compared against other classifications to provide well defined probabilities. This will additionally improve the statistical information of the classifier;

\item Including external secure classifications, like those based on {\it Gaia} proper motions \citep{gaia} or {\it Euclid} morphologies. We would need only to add a new boolean prior in Eq.~(\ref{pdf_pt}) to properly include this extra information in our PDFs.
\end{itemize}

These improvements will be explored in future versions of the classifier.

\section{Conclusions}\label{conclusions}
We have implemented a probabilistic morphological classification of the 251k J-PLUS EDR sources over 31.7 deg$^2$. Our Bayesian classifier is based on the distribution of J-PLUS sources in the concentration vs. magnitude space, where two populations are present: a compact sequence of point-like sources (stars) and a cloud of extended sources (galaxies). We modelled such distributions, including uncertainties, with a skewed Gaussian for compact objects and a log-normal function for the extended ones. The derived model and the number density prior based on J-PLUS EDR data were used to estimate the Bayesian probability of a source to be star or galaxy. This procedure was applied pointing-by-pointing to account for varying observing conditions and sky position, with stars being more numerous closer to the Milky Way disc. Finally, we combined the morphological information from $g$, $r$, and $i$ broad bands to have the morphological type PDF of each source.

We find that our PDF-based classifier provides a reliable boolean classification at $r \leq 20$, and a meaningful statistical classification at $r \leq 21$. This extra magnitude gained by Bayesian analysis was also estimated by \citet{scranton02} using SDSS dataset. The comparison with SDSS in the common areas is satisfactory up to $r \sim 21$, with consistent numbers of stars and galaxies, and consistent distributions in concentration and $(g-i)$ colour spaces. Future versions of the classifier will include colour information from the twelve photometric J-PLUS bands, improving the classification of low signal-to-noise sources. 

The derived probabilities were used to compute the pointing-by-pointing number counts of 150k stars and 101k galaxies in the J-PLUS EDR with $15 < r \leq 21$. The normalization of the stellar number counts increases as we approach to the Milky Way disc. Moreover, the higher latitude pointings seem to present a double power-law shape, that can be interpreted as the dominance of halo stars at $r \gtrsim 19$ in these fields \citep[e.g.][]{gao13}. Regarding galaxy number counts, our values are in good agreement with previous results in the literature. The pointing-to-pointing scatter increases at brighter magnitudes, reflecting the larger cosmic variance.

Using the stellar and galaxy number densities in the EDR as representative, we expect to detect $\sim40$ millions stars and $\sim25$ million galaxies at J-PLUS completion, significantly improving the study of the Milky Way halo structure and our knowledge about galaxy formation and evolution in the nearby Universe \citep{cenarro18}.

\begin{acknowledgements}
We dedicate this paper to the memory of our six IAC colleagues and friends who
met with a fatal accident in Piedra de los Cochinos, Tenerife, in February 2007,
with a special thanks to Maurizio Panniello, whose teachings of \texttt{python}
were so important for this paper.

Based on observations made with the JAST/T80 telescope at the Observatorio Astrof\'{\i}sico de 
Javalambre (OAJ), in Teruel, owned, managed and operated by the Centro de Estudios de F\'{\i}sica del 
Cosmos de Arag\'on. We acknowledge the OAJ Data Processing and Archiving Unit 
(UPAD) for reducing and calibrating the OAJ data used in this work.

Funding for the J-PLUS Project has been provided by
the Governments of Spain and Arag\'on through the Fondo de Inversiones
de Teruel, the Arag\'on Government through the Reseach Groups E96 and E103,
the Spanish Ministry of Economy and Competitiveness (MINECO; under
grants AYA2015-66211-C2-1-P, AYA2015-66211-C2-2, AYA2012-30789 and ICTS-2009-14),
and European FEDER funding (FCDD10-4E-867, FCDD13-4E-2685). 

B.~A. acknowledges received funding from the European Union’s Horizon 2020 research and innovation programme under the Marie Sklodowska-Curie grant agreement No. 656354.

V.~M.~P. and D.~D.~W. acknowledge partial support from grant PHY 14-30152; Physics Frontier Center/JINA Center for the Evolution of the Elements (JINA-CEE), awarded by the US National Science Foundation.

K.~V. acknowledges the {\it Juan de la Cierva - Incorporaci\'on} fellowship, IJCI-2014-21960, of the Spanish government.

Funding for the SDSS and SDSS-II has been provided by the Alfred P. Sloan Foundation, the Participating Institutions, the National Science Foundation, the U.S. Department of Energy, the National Aeronautics and Space Administration, the Japanese Monbukagakusho, the Max Planck Society, and the Higher Education Funding Council for England. The SDSS Web Site is \url{www.sdss.org}.

The SDSS is managed by the Astrophysical Research Consortium for the Participating Institutions. The Participating Institutions are the American Museum of Natural History, Astrophysical Institute Potsdam, University of Basel, University of Cambridge, Case Western Reserve University, University of Chicago, Drexel University, Fermilab, the Institute for Advanced Study, the Japan Participation Group, Johns Hopkins University, the Joint Institute for Nuclear Astrophysics, the Kavli Institute for Particle Astrophysics and Cosmology, the Korean Scientist Group, the Chinese Academy of Sciences (LAMOST), Los Alamos National Laboratory, the Max-Planck-Institute for Astronomy (MPIA), the Max-Planck-Institute for Astrophysics (MPA), New Mexico State University, Ohio State University, University of Pittsburgh, University of Portsmouth, Princeton University, the United States Naval Observatory, and the University of Washington.

This research made use of \texttt{Astropy}, a community-developed core \texttt{Python} package for Astronomy \citep{astropy}, and \texttt{Matplotlib}, a 2D graphics package used for \texttt{Python} for publication-quality image generation across user interfaces and operating systems \citep{pylab}.

\end{acknowledgements}

\bibliography{biblio}
\bibliographystyle{aa}

\end{document}